\documentclass[a4paper,11pt]{article}
\pdfoutput=1 

\usepackage{jheppub} 

\usepackage[T1]{fontenc} 

\usepackage{subfigure}	
\def\beq{\begin{equation}}
\def\eeq{\end{equation}}

\usepackage{graphicx}
\usepackage{amsmath}
\usepackage{amssymb}
\usepackage{amsthm}
\usepackage{tikz}
\usepackage{mathrsfs}
\usepackage{multirow}
\usepackage{scalerel}
\usepackage{mathtools}
\usepackage{textcomp}
\usepackage{dsfont}
\usepackage{blkarray}
\usepackage{bm}
\usepackage{todonotes}
\usepackage{hyperref}
\usepackage{pgf-pie}
\usepackage{slashed}
\usepackage{braket}
\usepackage[compat=1.1.0]{tikz-feynman}
\allowdisplaybreaks
\graphicspath{{Figures}}

\def\cF{\mathcal{F}}

\def\cM{\mathcal{M}}
\def\cN{\mathcal{N}}

\def\cS{\mathcal{S}}

\newcommand{\nn}{\nonumber}

\definecolor{darkyellow}{rgb}{0.5, 0.5, 0.0}
\definecolor{darkpurple}{rgb}{0.5, 0.2, 0.8}
\definecolor{darkblue}{rgb}{0.0, 0.0, 0.8}
\definecolor{darkgreen}{rgb}{0.0, 0.4, 0.0}
\definecolor{darkred}{rgb}{0.5, 0.0, 0.0}

\preprint{ZU-TH 51/23}
\title{Soft Theorem to Three Loops in QCD and ${\cal N} = 4$ Super Yang-Mills Theory}

\author[a]{Wen Chen,}
\emailAdd{chenwenphy@zju.edu.cn}
\author[b,a]{Ming-xing Luo,}
\emailAdd{mingxingluo@csrc.ac.cn}
\author[c]{Tong-Zhi Yang,}
\emailAdd{toyang@physik.uzh.ch}
\author[d]{and Hua Xing Zhu}
\emailAdd{zhuhx@pku.edu.cn}

\affiliation[a]{Zhejiang Institute of Modern Physics, School of Physics, Zhejiang University, Hangzhou, Zhejiang
310027, China}
\affiliation[b]{Complex Systems Division, Beijing Computational Science Research Center, Beijing, 100193,
China}
\affiliation[c]{Physik-Institut, Universit\"at Z\"urich, Winterthurerstrasse 190, CH-8057 Z\"urich, Switzerland}
\affiliation[d]{School of Physics, Peking University, Beijing, 100871, China}

\abstract{
The soft theorem states that scattering amplitude in gauge theory with a soft gauge-boson emission can be factorized into a hard scattering amplitude and a soft factor. In this paper, we present calculations of the soft factor for processes involving two hard colored partons, up to three loops in QCD. To accomplish this, we developed a systematic method for recursively calculating relevant Feynman integrals using the Feynman-Parameter representation. Our results constitute an important ingredient for the subtraction of infrared singularities at N$^4$LO in perturbative QCD. Using the principle of leading transcendentality between QCD and ${\cal N}=4$ super Yang-Mills theory, we determine the soft factor in the latter case to three loops with full-color dependence. As a by-product, we also obtain the finite constant $f_2^{(3)}$ in the Bern-Dixon-Smirnov ansatz analytically, which was previously known numerically only.
}

\begin{document} 

\maketitle
\flushbottom

\section{Introduction}
\label{sec:intro}

A remarkable property of gauge theories is that scattering amplitude containing a soft gauge boson can be factorized into a universal soft factor $\cS^\mu(k)$ and a hard scattering amplitude with the soft gauge boson removed,
\begin{equation}
\lim_{k\to 0} \cM_{n+1}(p_1, p_2, \cdots , p_n, k) = \cS_\mu(k)  \cM_n^\mu(p_1, p_2, \cdots, p_n) \,.\label{eq:softtheorem}
\end{equation}
This is known as the soft theorem~\cite{Low:1958sn,Yennie:1961ad,Weinberg:1965nx}. Note that for non-abelian gauge theory such as QCD, the soft factor is an operator acting on the color space of $n$-point amplitude. The soft theorem has found many applications in high energy physics both phenomenologically and theoretically.

In this paper we present a calculation of the soft theorem for soft gluon radiation from two hard partons to three loops in the perturbative QCD. Our main motivation is from the intimate relation between soft theorem and infrared behavior of QCD in higher order perturbation theory~\cite{Bassetto:1982ma,Bassetto:1983mvz,Berends:1988zn,Campbell:1997hg,Catani:1999ss}. In particular, the precision knowledge of the soft theorem in QCD is essential for constructing infrared subtraction terms in fixed order calculation~\cite{Catani:2007vq,Gehrmann-DeRidder:2005btv,Currie:2013vh,Czakon:2010td,Boughezal:2011jf,Boughezal:2015dva,Gaunt:2015pea,DelDuca:2016ily,Caola:2017dug,Bertolotti:2022aih}. It also contributes to the calculation of various soft function in Soft-Collinear Effective Theory~(SCET)~\cite{Bauer:2000ew,Bauer:2000yr,Bauer:2001yt,Bauer:2002nz,Beneke:2002ph}.
In this paper, we focus on the calculation of the soft factor with two hard-scattering partons and a single soft gluon emission. While it is not the most general soft factor beyond one-loop, it suffices for applications to some of the most important processes, such as the Drell-Yan process, $e^+e^-$ to dijet, and 1+1 jet production in deep-inelastic scattering. The one-loop contribution of it was calculated more than two decades ago~\cite{Giele:1991vf,Kunszt:1994np,Bern:1995ix, Catani:1996vz,Bern:1998sc,Bern:1999ry,Catani:2000pi}. The two-loop soft factor was initially extracted from the soft limit of splitting amplitude up to ${\cal O}(\epsilon^0)$ in dimensional regularization parameter~\cite{Badger:2004uk}, and was obtained through to ${\cal O}(\epsilon^2)$ and beyond, either by direct calculation in SCET~\cite{Li:2013lsa}, or by extracting from amplitude~\cite{Duhr:2013msa}. The two-loop soft factor constitutes an essential contribution to the total cross section of Higgs boson production at N$^3$LO in the threshold limit~\cite{Anastasiou:2013mca,Anastasiou:2014vaa,Li:2014afw,Li:2014bfa}. 
The two-loop soft factor is also an important ingredient for constructing infrared subtraction for generic perturbative QCD calculation at N$^3$LO. Besides the two-loop soft factor for single gluon emission, also relevant are the one-loop double-parton soft emission~\cite{Zhu:2020ftr,Catani:2021kcy,Czakon:2022dwk}, and the tree-level triple-parton soft emission~\cite{Catani:2019nqv,DelDuca:2022noh,Catani:2022hkb}. In addition, starting from two loops, a nontrivial color structure that correlates more than two partons first arises and has been computed in \cite{Dixon:2019lnw}. To further push the theoretical accuracy towards the N$^4$LO frontier for scattering cross section, in this paper we perform the calculation of the single soft-emission soft factor with two hard partons at three loops through ${\cal O}(\epsilon^2)$. 

To facilitate the calculation, we have developed a systematic approach to calculate single-scale soft integrals using Feynman parameter representation. Our main idea is to introduce an auxiliary scale in the parametric representation and directly construct differential equations in the parametric representation with respect to this scale. A parametric integral is nothing but a multi-fold integral. Thus, an auxiliary scale can trivially be introduced by leaving one fold of integration untouched. The obtained integrals can be calculated by using the standard differential-equation method~\cite{Kotikov:1990kg,Remiddi:1997ny}. The boundary conditions of the differential-equation system can again be expressed in terms of parametric integrals, which can be calculated by further applying this method. Thus, this method allows us to calculate Feynman integrals recursively until the boundary conditions can be trivially determined.

Besides phenomenological interests, the soft factor is also useful in determining quantities of theoretical interests. For instance, since the soft factor can be understood as the soft limit of the corresponding full amplitude, it shares the same iterative structure of the full amplitude in the maximally supersymmetric $\cN=4$ Yang-Mills theory (MSYM)~\cite{Anastasiou:2003kj,Bern:2005iz}. It was conjectured by Bern, Dixon, and Smirnov in ref.~\cite{Bern:2005iz} that the planar maximally helicity violating (MHV) amplitudes in MSYM can be obtained iteratively. Specifically, the $l$-loop planar MHV $n$-point amplitude in MSYM is determined by the one-loop amplitude up to some kinematic-independent constants, which are known to three loops numerically~\cite{Spradlin:2008uu}. Assuming the principle of transcendentality~\cite{Kotikov:2004er}, we obtain the soft factor in MSYM from reading off the leading transcendental part of the QCD results. We obtain the analytic expression for the three-loop constant $f_2^{(3)}$ in the BDS ansatz, which agrees well with the previously numerically determined one~\cite{Spradlin:2008uu}. In addition, we also predict the full-color dependence for the soft function in MSYM at three loops, which provides a test to the three-loop non-planar form factor of $1\to 3$ decay in MSYM~\cite{Lin:2021qol}, once the relevant master integrals there are computed.

The rest of this paper is organized as follows: in sec.~\ref{sec:CalcWeinbSoftTheorThrLoops}, we describe the method to calculate the soft factor based on an effective theory. The result is expressed in terms of single-scale soft master integrals. In sec.~\ref{sec:masterintCalculation}, we develop a systematic method to calculate these master integrals recursively based on the differential-equation method, with all the boundary integrals evaluated to gamma functions. The final results in both QCD and SYM are presented in sec.~\ref{sec:SoftTheorThrLoopsQCDSYM}.


\section{Calculation of QCD soft theorem to three loops}\label{sec:CalcWeinbSoftTheorThrLoops}

In this section we introduce the method for constructing the integrand for loop-level soft factor in QCD. Our approach is based on SCET, where the soft factor can be expressed as a transition matrix element of soft Wilson lines from vacuum to single gluon state. We use this definition to construct the integrand through three loops.

\subsection{Soft theorem from Soft-Collinear Effective Theory}

The soft theorem with an outgoing soft gluon can be well described by Wilson lines in Soft-Collinear Effective Theory~(SCET)~\cite{Bauer:2000ew,Bauer:2000yr,Bauer:2001yt,Bauer:2002nz,Beneke:2002ph} (See also, for example, ref.~\cite{Collins:2011zzd} for a discussion from QCD factorization). That is,
\begin{align}
\label{eq:softDefinition}
\varepsilon^\mu(q)  J_\mu(q) =  \braket{q| \int d^4x e^{i x \cdot q} \, \text{T} \bigg[ \prod_{k=1}^m Y_k(x) \bigg] |0}^{}, \,
\end{align}
where $Y_k(x)$ is a semi-infinity Wilson line standing for the color source of an external hard parton. In this paper, we restrict ourselves to the case of two Wilson lines with $m=2$. For an outgoing Wilson line, it starts from the origin and extends to null infinity, 
\begin{align}
\label{eq:Outsoftwilsonline}
    Y_k(x) = \text{P} \exp \left(  i g_s \int_0^{\infty} ds\, n_k \cdot A^a_s(x+ s n_k ) \mathbf{T}_k^a \right),
\end{align}
where the subscript '$s$' in $A_s^a$ refers to the soft gluon field. Similarly, an incoming Wilson line is defined as
\begin{align}
\label{eq:Insoftwilsonline}
    Y_k(x) =\text{P} \exp \left( i g_s \int_{-\infty}^{0} ds\, n_k \cdot A^a_s(x+ s n_k ) \mathbf{T}_k^a \right).
\end{align}
In the above equation, the P refers to path ordering
\begin{align}
    \text{P} \big[ \mathbf{A}(x+s n_k) \mathbf{A}& (x+t n_k) \big]  = \theta (s-t)  \mathbf{A} (x+s n_k) \mathbf{A}(x+t n_k) \nn
    \\ & +  \theta (t-s) \mathbf{A}(x+t n_k) \mathbf{A}(x+s n_k) \,, 
\end{align}
where we define $\mathbf{A}(x+ s n_k) = A^a(x+s n_k) \mathbf{T}_k^a$, and $\mathbf{T}_k^a$ is the color-charge operator defined in the color space formalism~\cite{Catani:1996vz}. For an outgoing quark (incoming anti-quark), $\left(\mathbf{T}_k^a\right)_{ij} = \left(t^a\right)_{ij}$, for an outgoing anti-quark (incoming quark), $\left(\mathbf{T}_k^a\right)_{ij} = -\left(t^a\right)_{ji}$, for a gluon, $\left(\mathbf{T}_k^a\right)_{bc} = -i f^{abc}$, where $f^{abc}$ are structure constants, and $t^a$ are the Gell-Mann matrices with the normalization $\text{Tr} [ t^a t^b]  = \frac{1}{2} \delta^{ab}$.

According to Lorentz and color structures, the soft factor $J_\mu$ with two Wilson lines can be decomposed into the following form up to three loops:
\begin{align}
\label{eq:ColorDecomposition}
    J_\mu^a(q) =&-\frac{g_s}{2}\left( \frac{n^\mu_1}{n_1 \cdot q} - \frac{n^\mu_2}{n_2 \cdot q} \right) \bigg[ \left(\mathbf{T}^{a}_{1} - \mathbf{T}^{a}_{2}\right)  +  2 i f^{a b c}\left( \mathbf{T}^{b}_{1} \mathbf{T}^{c}_2 - \mathbf{T}^{b}_{2} \mathbf{T}^{c}_1 \right) \, B_{12}  \nonumber
    \\
    &-  \left( \mathbf{T}^{b}_{1} \mathbf{T}^{c}_1 \mathbf{T}^{d}_2 - \mathbf{T}^{b}_{2} \mathbf{T}^{c}_2 \mathbf{T}^{d}_1 \right) \left( C_{12} \,  d_A^{a b c d}  + D_{12} \,d_F^{a b c d} N_f \right)  \bigg] + \mathcal{O}(\alpha_s^4)\,,
\end{align}
where $n_1^2 = n_2^2 = 0$ are two light-like vectors. The form factor $B_{12}$ starts to contribute at one loop. The quadrupole invariant tensor $d_A^{abcd}$ and $d_F^{abcd}$ are defined by 
\begin{align}
    d_R^{abcd} = \frac{1}{24} \text{Tr}\big[ \mathbf{T}^a \mathbf{T}^b \mathbf{T}^c \mathbf{T}^d \big]_R + \text{symmetric permutations}\,,
\end{align}
and their coefficients $C_{12},\, D_{12}$ only receive contributions starting from three-loop order. We stress that all these scalar factors don't depend on the particular representation of the Wilson lines, a form of Casimir scaling. The form of eq.~\eqref{eq:ColorDecomposition} is constructed from scaling symmetry and dimensional analysis. It has been checked by an explicit computation which will be described in detail below. 

A related quantity, the $l$-loop Eikonal function can be derived from the soft factor, 
\begin{align}
\label{eq:eikonalfunction}
    S_{12}^{(l)}(q) = \frac{1}{4 N_R C_R}  \text{Tr}  \bigg\{ \left[ \varepsilon^{\mu} J_\mu^{a(l)} \right]  \left[ \varepsilon^{\nu} J_{\nu}^{a (0)} \right]^*(q)\bigg\}\,, 
\end{align}
where in SU($N_c$) group $N_F = N_c,\, C_F =  (N_c^2-1)/(2 N_c)$ and $N_A = N_c^2-1,\, C_A= N_c$ for quarks and gluons respectively. The eikonal function directly contributed to the soft-virtual cross section of Drell-Yan or Higgs production, see e.g. \cite{Anastasiou:2014vaa,Li:2014afw}. Here and in the following, we always expand the quantity in
\begin{align}
\label{eq:alsexpansion}
a_s = \frac{\alpha_s}{4 \pi}    
\end{align}
with $\alpha_s = g_s^2/(4 \pi)$, for example,
\begin{align}
    S_{12}(q) = (4 \pi)^2 a_s \sum_{l=0}^\infty a_s^l\,  S_{12}^{(l)}(q) \,.
\end{align}
The $S_{12}^{(0)}$ is the well-known tree-level Eikonal function, 
\begin{align}
\label{eq:treeEikonal}
    S_{12}^{(0)} = \frac{n_1 \cdot n_2}{ 2 n_1 \cdot q \, n_2 \cdot q}\,.
\end{align}
We are mainly interested in the higher-order corrections for scalar form factors in eq.~\eqref{eq:ColorDecomposition} and the tree-level Eikonal function in eq.~\eqref{eq:treeEikonal}. The soft factor in general depends on the direction of the Wilson lines (incoming or outgoing). However, due to a rescaling symmetry $n_1^\mu \to \lambda_1 n_1^\mu$ and $n_2^\mu \to \lambda_2 n_2^\mu$ for arbitrary $\lambda_1$ and $\lambda_2$, this dependence can be fully encoded in terms of a factor $S_\epsilon$:
\begin{align}
    S_{\epsilon} =\left( 4 \pi S_{12}^{(0)} \mu^2 e^{-\gamma_E} \frac{e^{-i \lambda_{12} \pi }}{ e^{-i \lambda_{1q} \pi }  e^{-i \lambda_{2q} \pi} } \right)^\epsilon \,,
\end{align}
where $\epsilon= (4-d)/2$ is the dimensional regulator, $\lambda_{AB}$ in the phase factor $e^{-i \lambda_{AB} \pi} $ is 1 if A and B are both incoming or outgoing, and $\lambda_{AB}=0$ for other cases (see for example~\cite{Catani:2000pi}). By factoring out the dependence on $S_\epsilon$ for eq.~\eqref{eq:ColorDecomposition} and eq.~\eqref{eq:eikonalfunction} at each order, the remaining contributions are not sensitive to the direction of Wilson lines,
\begin{align}
\label{eq:factorOutSe}
    S_{12}^{(l)}(q) &= S_{12}^{(0)}(q) S_\epsilon^l \,r^{(l)}_{12}\,,
    \nonumber
    \\
    B_{12}^{(l)} =  S_\epsilon^l \, b_{12}^{(l)}\,,  C_{12}^{(l)} &=  S_\epsilon^l \, c_{12}^{(l)}\,, D_{12}^{(l)} =  S_\epsilon^l \, d_{12}^{(l)} \,. 
\end{align}
In this paper, we determine $r_{12}$ and $b_{12},\, c_{12}, \, d_{12}$ in above equation to three-loop order.

\subsection{Construction of loop integrand for Soft theorem}
To construct the loop integrand for the soft theorem, we first derive the effective Feynman rules for soft Wilson lines as shown in eq.~\eqref{eq:Outsoftwilsonline} and eq.~\eqref{eq:Insoftwilsonline}. It can be conveniently done by expanding the Wilson lines order by order in $g_s$. We get the following eikonal Feynman rules up to three gluon emissions (We checked explicitly that the Feynman rules with more gluon emissions are not needed for the computation of three-loop soft theorem), 
\begin{align}
\label{eq:nFeynmanRules}
    &\includegraphics[scale=1.7]{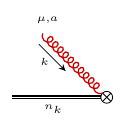}  \nonumber
    \\
    &\qquad  \to  \frac{-g_s n_k^\mu \,\mathbf{T}_k^a}{-n_k \cdot k+ i \delta_k \, 0^+}  \,, \nonumber
    \\
    &\includegraphics[scale=1.7]{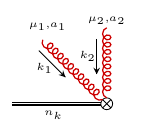}  \nonumber
    \\
    &\qquad  \to \frac{  g_s^2  n_k^{\mu_1} n_k^{\mu_2} }{-n_k \cdot \left( k_1 +k_2\right)+ i \delta_k \, 0^+}\left[  \frac{\mathbf{T}^{a_1}_k \mathbf{T}^{a_2}_k}{ - n_k \cdot k_1 + i 
 \delta_k\, 0^+}  +  \frac{\mathbf{T}^{a_2}_k \mathbf{T}^{a_1}_k }{ - n_k \cdot k_2 + i \delta_k \, 0^+}  \right]  \,, \nonumber
    \\
    &\includegraphics[scale=1.7]{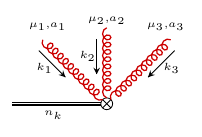} \nonumber
    \\
    & \qquad \to  \frac{-g_s^3 n_k^{\mu_1} n_k^{\mu_2} n_k^{\mu_3} }{  -n_k \cdot \left( k_1 +k_2+k_3\right)  +i \delta_k \, 0^+} \bigg[ \frac{\mathbf{T}^{a_1}_k \mathbf{T}^{a_2}_k \mathbf{T}^{a_3}_k  }{ -n_k \cdot (k_1+k_2) +i \delta_k\, 0^+ }  \frac{1}{   -n_k \cdot k_1    + i \delta_k \, 0^+}  \nonumber 
    \\
    & \qquad \qquad \qquad \qquad  \qquad \qquad \qquad \qquad \qquad+ \textit{permutations}\bigg] \,, 
\end{align}
where the sign of Feynman's prescription $i \delta_k\, 0^+$ stems from the path along the light cone of outgoing or incoming Wilson lines. The $\delta_k$ is 1 for an outgoing Wilson line and $\delta_k =-1$ for an incoming Wilson line. The color-charge operator $\mathbf{T}^a_k$ is the same operator as defined in eq.~\eqref{eq:Outsoftwilsonline} and eq.~\eqref{eq:Insoftwilsonline}. In the above equations, plus \textit{permutations} indicates a summation over all external gluon indices (simultaneous permutation of $\mu_i,a_i,k_i$). Each vertex for the Wilson line above can be understood as a sum of conventional diagrams with soft gluon emissions. For example, the vertex with two soft gluons in eq.~(\ref{eq:nFeynmanRules}) for an outgoing Wilson line can be understood as the sum of the following two diagrams:\\

\begin{tikzpicture}
  \begin{feynman}
    \vertex (a1) ;
    \vertex at ($(a1) + (1.6cm, 0.8cm)$) (b1) ;
    \vertex at ($(a1) + (3.2cm, 1.6cm)$) (c1) ;
    \vertex at ($(a1) + (4.8cm, 2.4cm)$) (d1) {$n_k$};
    \vertex at ($(d1) + (0cm, -3.2cm)$) (e1) ;
    \vertex at ($(d1) + (0cm, -1.6cm)$) (f1) ;
    \vertex at ($(a1) + (8cm, 0cm)$) (a2) ;
    \vertex at ($(b1) + (8cm, 0cm)$) (b2) ;
    \vertex at ($(c1) + (8cm, 0cm)$) (c2) ;
    \vertex at ($(d1) + (8cm, 0cm)$) (d2) ;
    \vertex at ($(e1) + (8cm, 0cm)$) (e2) ;
    \vertex at ($(f1) + (8cm, 0cm)$) (f2) ;
    \vertex at ($(a1) + (7cm, 1cm)$) (a) {$+$};
    \diagram* {
    (a1) --[double,momentum=$-k_1-k_2$] (b1) -- [double,with arrow=0.5,momentum=$-k_1$] (c1) -- [double] (d1),
    (e1) --[gluon,momentum=$k_2$] (b1),
    (f1) --[gluon,momentum=$k_1$] (c1),
    };
    \diagram* {
    (a2) --[double] (b2) -- [double,with arrow=0.5] (c2) -- [double] (d2),
    (f2) --[gluon] (b2),
    (e2) --[gluon] (c2), 
    }; 
  \end{feynman}
\end{tikzpicture}~~.\\
Here the effective coupling between a soft gluon and an eikonal line is $ig_s \mathbf{T}_k ^a n_k^\mu$. And the propagator of the eikonal line with momentum $k$ is $\frac{i}{n_k\cdot k+i0^+}$.

\begin{figure}
\begin{center}
\begin{minipage}{4.6cm}
\includegraphics[scale=1.4]{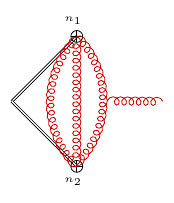} 
\end{minipage}
\begin{minipage}{4.6cm}
\includegraphics[scale=1.4]{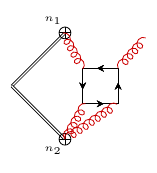} 
\end{minipage}
\begin{minipage}{4.6cm}
\includegraphics[scale=1.4]{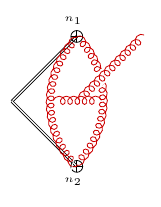} 
\end{minipage}
    \end{center}
    \caption{Sample Feynman diagrams for three-loop soft theorem. From left to right, when contracting with tree diagrams with single gluon emission, the first diagram is zero, the second diagram contributes to $d_R^{abcd} d_F^{abcd}$, and the third diagram contributes to sub-leading color only.}
    \label{figure:sampleDiagram}
\end{figure}

We generated in \texttt{QGRAF}~\cite{Nogueira:1991ex} all relevant Feynman diagrams, which implement particle interactions from standard QCD and interactions due to the above effective vertices with up to three-gluon emissions. In figure~\ref{figure:sampleDiagram}, we show some sample Feynman diagrams. The amplitude is invariant under the rescaling of $n_1,\,n_2$, such that the only scale is $\mu^2 S_{12}^{(0)}$. Therefore, the soft factor only receives contributions from one-particle-irreducible (1PI) diagrams with a number 550 at three loops. Subsequently, an in-house \texttt{Mathematica} code was used to substitute the Feynman rules into Feynman diagrams, and \texttt{FORM}~\cite{Vermaseren:2000nd,Kuipers:2012rf,Ruijl:2017dtg} and \texttt{Color.h}~\cite{vanRitbergen:1998pn} were used to evaluate Dirac and color algebra. To verify the (generalized) Casimir scaling principle, we use the effective Feynman rules as shown in eq.~\eqref{eq:nFeynmanRules} for Wilson lines in both fundamental and adjoint representations. 
Regarding the topology classification, the package \texttt{Apart}~\cite{Feng:2012iq} was first used to eliminate the linear dependence of propagators largely due to the multiple linear propagators from the effective vertices. After the partial fraction, we found 780 topologies that were then reduced to 160 topologies by applying a self-written code. The code implemented a simple algorithm that tries to find a loop momentum transformation relating the two topologies with each other by carefully searching all possible loop momentum transformations. We noted that a similar algorithm was also implemented in the public package \texttt{Reduze 2}~\cite{vonManteuffel:2012np}~\footnote{We thank Andreas von Manteuffel for pointing this out to us.}. By appending some proper propagators stemming from irreducible numerators, the 160 topologies can be further mapped into 25 integral families. The definition of these integral families can be found in Appendix.~\ref{sec:integralsFamilies}.

The integration-by-parts (IBP)~\cite{Chetyrkin:1981qh} reductions were done by \texttt{Kira}~\cite{Klappert:2020nbg} equipped with \texttt{FireFly}~\cite{Klappert:2019emp}, which implements the Laporta algorithm~\cite{Laporta:2000dsw} as well as finite fields and function reconstruction techniques~\cite{vonManteuffel:2014ixa,Peraro:2016wsq}. After IBP reductions, we found 52 master integrals and only 49 of them appeared in the amplitude, and the master integrals were found to appear only in six integral families.

To check the gauge invariance of the amplitude, we use the Feynman gauge as well as the light cone gauge for the polarization summation of the external gluon as shown in eq.~\eqref{eq:eikonalfunction},
\begin{align}
    \varepsilon_\mu(q)  \varepsilon^*_\nu(q) = - g_{\mu \nu} + \sigma \frac{n_\mu q_\nu + n_\nu q_\mu}{ n \cdot q} , \quad \sigma =0 \text{ or } 1 \,.
\end{align}
For internal gluons, we use the $R_\xi$ gauge but with the amplitude truncating at $(1-\xi)^1$, 
\begin{align}
    D_{\mu \nu}(l) = \frac{i}{l^2} \left[ - g_{\mu \nu}  +  (1-\xi) \frac{l_\mu l_\nu}{l^2} \right] \,.
\end{align}
We found the amplitude is indeed gauge invariant provided that six extra relations exist within the 49 master integrals. We verified explicitly these six extra relations by computing all 49 master integrals as shown in section~\ref{sec:masterintCalculation}.


\section{Calculation of master integrals}
\label{sec:masterintCalculation}

In this section, we present the details of our approach to compute the single-scale soft master integrals. Our approach is based on Feynman parameter representation. Using the differential equation with respect to an auxiliary scale appearing in the intermediate step and reduction of integrals in Feynman parameter representation, we manage to compute all master integrals iteratively, with the final boundary conditions coming from simple Gamma functions.

\subsection{Differential equations}\label{subsec:DiffEquat}

We calculate the master integrals by using the differential-equation method~\cite{Kotikov:1990kg,Remiddi:1997ny}. For soft integrals, the scale dependence is trivial. To get a nontrivial differential-equation system, we need to introduce an auxiliary scale. While there are several widely used methods to achieve this, such as the Drinfeld-associator method~\cite{Henn:2013nsa} and the auxiliary-mass-flow method~\cite{Liu:2017jxz,Liu:2022mfb,Liu:2022tji,Liu:2022chg}, these methods are less appropriate for the calculation in this work. Because an extra mass scale on either an external line or an internal line may highly increase the complexity of the IBP reduction. A better choice is to introduce a scale that is essential to the integral to be calculated. It is evident that an extra scale can be introduced by leaving one fold of integration untouched, or equivalently, by inserting a delta function. For phase-space integrals, this can be done in the momentum space (see e.g. refs.~\cite{Zhu:2014fma,Chen:2020dpk}). While for normal loop integrals, an extra delta function in the momentum space may even complicate the calculation. A better choice is to insert a delta function in the parametric representation. The obtained integrals can further be reduced by using the method developed in refs.~\cite{Chen:2019mqc,Chen:2019fzm,Chen:2020wsh}.

We consider the calculation of the parametric integrals of the following form
\begin{equation}\label{Eq:ParInt}
\begin{split}
I(\lambda_0,\lambda_1,\ldots,\lambda_n)=&\frac{\Gamma(-\lambda_0)}{\prod_{i=m+1}^{n+1}\Gamma(\lambda_i+1)}\int \mathrm{d}\Pi^{(n+1)}\mathcal{F}^{\lambda_0}\prod_{i=1}^{m}x_i^{-\lambda_i-1}\prod_{i=m+1}^{n+1}x_i^{\lambda_i}\\
\equiv&\int \mathrm{d}\Pi^{(n+1)}\mathcal{I}^{(-n-1)}\,.
\end{split}
\end{equation}
Here the integration measure is $\mathrm{d}\Pi^{(n)}\equiv\prod_{i=1}^{n+1}\mathrm{d}x_i\delta(1-E^{(1)}(x))$, with $E^{(n)}(x)$ a positive definite homogeneous function of $x$ of degree $n$. The region of integration for $x_i$ is $(0,~\infty)$ when $i>m$ and $(-\infty,~\infty)$ when $i\leqslant m$. $\mathcal{F}$ is a homogeneous polynomial of $x$ of degree $L+1$. For integrals with momentum-space correspondences, $L$ is the number of loops. For loop integrals, the polynomial $\mathcal{F}$ is related to the well-known Symanzik polynomials $U$ and $F$ through $\mathcal{F}=F+Ux_{n+1}$. But here we consider the more general parametric integrals which may not have momentum-space correspondences. This generalization is necessary, because some asymptotically expanded integrals may not have any momentum-space correspondence~\cite{Beneke:1997zp,Pak:2010pt}. By virtue of the homogeneity of the integrands, it can be shown that the parametric integrals satisfy the equations
\begin{subequations}\label{eq:IBP1}
\begin{align}
0=&\int \mathrm{d}\Pi^{(n+1)}\frac{\partial}{\partial x_i}\mathcal{I}^{(-n)},&& i=1, 2,\ldots, m,\\
0=&\int \mathrm{d}\Pi^{(n+1)}\frac{\partial}{\partial x_i}\mathcal{I}^{(-n)}+\delta_{\lambda_i0}\int \mathrm{d}\Pi^{(n)}\left.\mathcal{I}^{(-n)}\right|_{x_i=0},&& i=m+1, m+2,\ldots, n+1.
\end{align}
\end{subequations}

A parametric integral can be understood as a function of the indices $\lambda_i$. Then we can define the following operators.
\begin{align*}
\mathcal{R}_iI(\lambda_0,\dots,\lambda_i,\dots,\lambda_n)=&(\lambda_i+1)I(\lambda_0,\dots,\lambda_i+1,\dots,\lambda_n),\\
\mathcal{D}_iI(\lambda_0,\dots,\lambda_i,\dots,\lambda_n)=&I(\lambda_0,\dots,\lambda_i-1,\dots,\lambda_n),\\
\mathcal{A}_iI(\lambda_0,\dots,\lambda_i,\dots,\lambda_n)=&\lambda_iI(\lambda_0,\dots,\lambda_i,\dots,\lambda_n).
\end{align*}
It is understood that
\begin{equation*}
I(\lambda_0,\dots,\lambda_{i-1},-1,\dots,\lambda_n)\equiv\int \mathrm{d}\Pi^{(n)}\left.\mathcal{I}^{(-n)}\right|_{x_i=0},\quad i=m+1,~m+2,~\cdots,~n.
\end{equation*}
We further define
\begin{align*}
\hat{x}_i=&\left\{
\begin{matrix}
\mathcal{D}_i&,&i=1,~2,\ldots,~m,\\
\mathcal{R}_i&,&i=m+1,~m+2,\ldots,~n+1,
\end{matrix}
\right.\\
\hat{z}_i=&\left\{
\begin{matrix}
-\mathcal{R}_i&,&i=1,~2,\ldots,~m,\\
\mathcal{D}_i&,&i=m+1,~m+2,\ldots,~n+1,
\end{matrix}
\right.\\
\hat{a}_i=&\left\{
\begin{matrix}
-\mathcal{A}_i-1&,&i=1,~2,\ldots,~m,\\
\mathcal{A}_i&,&i=m+1,~m+2,\ldots,~n+1.
\end{matrix}
\right.
\end{align*}
And we formally define operators $\hat{z}_{n+1}$ and $\hat{x}_{n+1}$, such that $\hat{z}_{n+1}I=I$, and $\hat{x}_{n+1}^iI=\prod_{j=1}^i(\hat{a}_{n+1}+j)I$, with $\hat{a}_{n+1}=-(L+1)\hat{a}_0-\sum_{i=1}^n(\hat{a}_i+1)-1$. We assume that $\hat{x}_{n+1}$ is always to the right of $U(\hat{x})$ in $\mathcal{F}(\hat{x})$. By using these operators, we can write eq.~(\ref{eq:IBP1}) in the following form~\footnote{Notice that here the definition of $\hat{a}_{n+1}$ is slightly different from that in refs.~\cite{Chen:2019fzm,Chen:2020wsh}. The reason is that, with the definition of $\hat{a}_{n+1}$ in refs.~\cite{Chen:2019fzm,Chen:2020wsh}, eq.~(\ref{eq:DE}) is invalid if the first Symanzik polynomial $U$ depends on $y$. And in this paper we do need to consider the situation where $U$ depends on $y$. With the new definition, there is an extra $\hat{x}_{n+1}$ in eq.~(\ref{eq:IBP2}) (compared with eq.~(2.6) in ref.~\cite{Chen:2020wsh}). And eq.~(2.11) in ref.~\cite{Chen:2019fzm} becomes $D_0\mathcal{F}+\mathcal{A}_0\approx0$. The definition of $\hat{x}_{n+1}$ is only of formal sense, and $\hat{x}_{n+1}^i\hat{x}_{n+1}^jI$ should be understood as $\hat{x}_{n+1}^{i+j}I$ rather than $\hat{x}_{n+1}^i\left(\hat{x}_{n+1}^jI\right)$. In practical calculations, we always express $\hat{x}_{n+1}^i$ in terms of $\hat{a}_{n+1}$ from the very beginning.}:
\begin{equation}\label{eq:IBP2}
    \left[\mathcal{D}_0\frac{\partial\mathcal{F}(\hat{x})}{\partial\hat{x}_i}-\hat{z}_i\right]\hat{x}_{n+1}I=0.
\end{equation}

Let $y$ be a kinematical variable, then it is easy to see that the parametric integrals satisfy the equation
\begin{equation}\label{eq:DE}
    \frac{\partial}{\partial y}I=-\mathcal{D}_0\frac{\partial\mathcal{F}}{\partial y}I.
\end{equation}

To get a nontrivial scale dependence, we insert a delta function into the parametric integral in eq.~(\ref{Eq:ParInt}), and get
\begin{equation}\label{eq:Inserty}
\begin{split}
    I(\lambda_0,\lambda_1,\ldots,\lambda_n)=&\int \mathrm{d}\Pi^{(n+1)}\mathrm{d}y~\delta(y-E^{(0)}(x))\mathcal{I}^{(-n-1)}.
\end{split}
\end{equation}
Here the function $E^{(n)}$ is the one defined below eq.~(\ref{Eq:ParInt}). Equation~(\ref{eq:IBP1}) also holds for eq.~(\ref{eq:Inserty}), since it is a consequence of the homogeneity of the integrand. In practical calculations, by a proper choice of the $E^{(0)}(x)$, we can eliminate one fold of integration by using the delta function. The resulting $y$-dependent integral is still of the form in eq.~(\ref{Eq:ParInt}). Thus it can be reduced in the parametric representation and then calculated by using the differential-equation method. Compared with the original integral, the $y$-dependent integral is one less fold of integration. By successive applications of this method, we can calculate Feynman integrals recursively.

The method described in this subsection applies to both normal loop integrals and phase-space integrals. But we do not consider phase-space integrals hereafter. That is, we take $m=0$ in eq.~(\ref{Eq:ParInt}).

\subsection{Rules for choosing $E^{(0)}$}
In principle, the function $E^{(0)}$ in eq.~(\ref{eq:Inserty}) can be chosen arbitrarily. Nevertheless, for a general choice of $E^{(0)}$, it is not easy to express the right-hand sides of eqs.~(\ref{eq:IBP1}) in terms of regular parametric integrals. In practical calculations, we choose $E^{(0)}$ to be of the form
\begin{equation}
    E^{(0)}=\frac{x_i}{x_j}.
\end{equation}
Then we can eliminate the integration with respect to $x_i$ by using the delta function. That is,
\begin{equation}
\begin{split}
    I(\lambda_0,\lambda_1,\ldots,\lambda_n)=&\int \mathrm{d}\Pi^{(n+1)}\mathrm{d}y~\delta(y-\frac{x_i}{x_j})\mathcal{I}^{(-n-1)}\\
    =&\int\mathrm{d}y\int\mathrm{d}\Pi^{(n)}~x_j\left.\mathcal{I}^{(-n-1)}\right|_{x_i=yx_j}\\
    \equiv&\frac{\Gamma(\lambda_i+\lambda_j+2)}{\Gamma(\lambda_i+1)\Gamma(\lambda_j+1)}\int\mathrm{d}y~y^{\lambda_i}I_y.
\end{split}
\end{equation}
For integrals that have momentum-space correspondences, this is equivalent to the method of combining two propagators with a Feynman parameter~\cite{Hidding:2022ycg}. The pair $\{x_i,x_j\}$ can still be arbitrarily chosen. A good choice may greatly simplify the calculation. In this section, we provide a method to choose $E^{(0)}$ wisely such that the number of regions of the asymptotic expansion for the obtained $y$-dependent integral is minimized. Consequently, the boundary conditions of the differential equations are simplified.

A general $\mathcal{F}$ polynomial is of the structure
\begin{equation}\label{eq:FPol}
\mathcal{F}=\sum_{a=1}^{A}\left(C_{\mathcal{F},a}\prod_i^{n+1}x_i^{\Lambda_{ai}}\right),
\end{equation}
where $C_{\mathcal{F},a}$ are some $x$-independent constants. This polynomial may not depend on any physical scale. Thus it does not make sense to talk about asymptotic expansion for the corresponding parametric integrals. Nevertheless, we can still formally introduce the notion of "region" for this polynomial by using the idea of the convex hull described in ref.~\cite{Pak:2010pt}. Specifically, a region $r$ is associated with a subset $S_r$ of $\{1,~2,\cdots,A\}$ and a $n+2$ dimensional vector $\bm{k}_r$, such that the number of elements of $S_r$ is not less than $n+1$, and
\begin{subequations}\label{eq:RegVect}
    \begin{align}
        \sum_{k=1}^{n+1}\Lambda_{ak}k_{r,k}=&k_{r,0},\quad a\in S_r,\\
        \sum_{k=1}^{n+1}\Lambda_{ak}k_{r,k}>&k_{r,0},\quad a\notin S_r.
    \end{align}
\end{subequations}
It is easy to see that $\Lambda_{ai}$ with $a\notin S_r$ is linearly independent of $\Lambda_{ai}$ with $a\in S_r$. Since $\mathcal{F}$ is a homogeneous polynomial of degree $L+1$, we have $\sum_{i=1}^{n+1}\Lambda_{ai}=L+1$. Thus, if $\Lambda_{bi}=\sum_{a\in S_r}c_{ba}\Lambda_{ai}$, we have $\sum_ac_{ba}=1$, and $\sum_i\Lambda_{bi}k_{r,i}=k_{r,0}\sum_ac_{ba}=k_{r,0}$. Hence $b\in S_r$. And it is easy to see that the cardinal number of $S_r$ should be smaller than $A$, because otherwise the corresponding parametric integral is scaleless. To see this, without loss of generality, we assume that $k_{r,1}\neq0$. Then we rescale $x_i$ with $i>1$ by $x_i\to x_ix_1^{k_{r,i}/k_{r,1}}$. If $S_r=\{1,~2,\dots,A\}$, the $x_1$ dependence of $\cF$ can be factored out. Thus the integration with respect to $x_1$ is scaleless.

Presently we will show that those regions defined by eqs.~(\ref{eq:RegVect}) are intimately related to the regions of the $y$-dependent integrals.

We consider the integrals obtained by replacing $x_i$ with $yx_j$. The corresponding $\mathcal{F}$ polynomial reads
\begin{equation}
    \mathcal{F}^\prime=\left.\mathcal{F}\right|_{x_i=yx_j}.
\end{equation}
For simplicity, we formally denote $y$ by $x_i$ for $\mathcal{F}^\prime$. Obviously, we have
\begin{equation}
    \mathcal{F}^\prime=\sum_{a=1}^{A}\left(C_{\mathcal{F},a}\prod_k^{n+1}x_k^{\Lambda^\prime_{ak}}\right),
\end{equation}
with
\begin{subequations}
    \begin{align}
        \Lambda^\prime_{aj}=&\Lambda_{aj}+\Lambda_{ai},\\
        \Lambda^\prime_{ak}=&\Lambda_{ak},\qquad k\neq j.
    \end{align}
\end{subequations}
It is easy to see that
\begin{subequations}
    \begin{align}
        \sum_{k=1}^{n+1}\Lambda^\prime_{ak}k^\prime_{r,k}=&k_{r,0},\quad a\in S_r,\\
        \sum_{k=1}^{n+1}\Lambda^\prime_{ak}k^\prime_{r,k}>&k_{r,0},\quad a\notin S_r,
    \end{align}
\end{subequations}
with
\begin{subequations}
    \begin{align}
        k^\prime_{r,i}\equiv&k_{r,i}-k_{r,j},\\
        k^\prime_{r,k}\equiv&k_{r,k},\qquad k\neq i.
    \end{align}
\end{subequations}

According to the convex-hull algorithm described in ref.~\cite{Pak:2010pt}, a vector $\bm{k}^\prime_r$ with $k^\prime_{r,i}>0$ gives exactly a region of asymptotic expansion in the limit of $y\to0$, because terms $\prod_i^{n+1}x_i^{\Lambda^\prime_{ai}}$ with $a\in S_r$ dominate $\mathcal{F}^\prime$ when $x_k$ scales as $x_k\sim y^{k^\prime_{r,k}}$. We denote
\begin{equation}\label{eq:defRij}
    R_{ij}=\left\{r|k_{r,i}>k_{r,j}\right\}.
\end{equation}
Since $k^\prime_{r,i}=k_{r,i}-k_{r,j}$, by choosing the pair $\{i,~j\}$ such that the cardinal number of the set $R_{ij}$ is minimized, the number of regions of asymptotic expansion is minimized.

Obviously, after expanding the $\mathcal{F}$ polynomial asymptotically in a region $r$, only terms in $S_r$ survive. Thus, by choosing the pair $\{i~,j\}$ such that the cardinal number of $S_r$  (denoted by $N_r$) is minimized, the boundary integrals are simplified.

As a summary, we choose the pair $\{i,~j\}$ according to the following rules:
\begin{itemize}
    \item [(1)] We choose the pair $\{i,~j\}$ such that the cardinal number of $R_{ij}$ is minimized, where $R_{ij}$ is defined in eq.~(\ref{eq:defRij}).
    \item[(2)] Among all the pairs satisfying the first rule, we choose the one such that $\max\{N_r|r\in R_{ij}\}$ is minimized, where $N_r$ is the cardinal number of $S_r$.
\end{itemize}

\subsection{Boundary integrals}

By using the method described in the previous sections, we can construct differential equations for the parametric integrals. The boundaries of the solutions of the differential equations can further be expressed in terms of parametric integrals. Thus, this algorithm can be carried out recursively. The algorithm terminates when the $\mathcal{F}$ polynomial has exactly $n+1$ monomials. In this case, the parametric integral can be expressed in terms of gamma functions. We have
\begin{equation}\label{eq:BoundInt}
\begin{split}
I(\lambda_0,\lambda_1,\ldots,\lambda_n)=&\frac{\Gamma(-\lambda_0)}{\prod_{i=1}^{n+1}\Gamma(\lambda_i+1)}\int \mathrm{d}\Pi^{(n+1)}\mathcal{F}^{\lambda_0}\prod_{i=1}^{n+1}x_i^{\lambda_i}\\
=&\frac{(L+1)\prod_{a=1}^{n+1}\left[\Gamma(\bar{\lambda}_a)C_{\mathcal{F},a}^{-\bar{\lambda}_a}\right]}{\parallel\Lambda\parallel\prod_{i=1}^{n+1}\Gamma(\lambda_i+1)}.
\end{split}
\end{equation}
with
\begin{equation}
    \bar{\lambda}_a=\sum_{i=1}^{n+1}(\Lambda^{-1})_{ia}(\lambda_i+1).
\end{equation}
Here $\Lambda_{ai}$ and $C_{\mathcal{F},a}$ are defined in eq.~(\ref{eq:FPol}), and $L$ is defined in the paragraph below eq.~(\ref{Eq:ParInt}).

The derivation of eq.~(\ref{eq:BoundInt}) is as follows.

We introduce a new set of variables
\begin{equation}
    u_a\equiv\prod_{i=1}^{n+1}x^{\Lambda_{ai}}.
\end{equation}
The Jacobian is
\begin{equation}
        \left|\left|\frac{\partial u_a}{\partial x_i}\right|\right|=\left|\left|\frac{u_a}{x_i}\Lambda_{ai}\right|\right|
        =\frac{\prod_{a=1}^{n+1}u_a}{\prod_{i=1}^{n+1}x_i}\parallel\Lambda\parallel
\end{equation}
For the integration measure $\mathrm{d}\Pi^{(n+1)}$, we choose $E^{(1)}=u_{n+1}^{\frac{1}{L+1}}$. Then we have
\begin{equation}
\begin{split}
I(\lambda_0,\lambda_1,\ldots,\lambda_n)=&\frac{(L+1)\Gamma(-\lambda_0)}{\parallel\Lambda\parallel\prod_{i=1}^{n+1}\Gamma(\lambda_i+1)}\int \prod_{a=1}^n\mathrm{d}u_a\\
&\left(C_{\mathcal{F},n+1}+\sum_{a=1}^{n}C_{\mathcal{F},a}u_a\right)^{\lambda_0}\prod_{a=1}^{n}u_a^{\sum_{i=1}^{n+1}\left(\Lambda^{-1}\right)_{ia}(\lambda_i+1)-1}\\
=&\frac{(L+1)\Gamma(-\lambda_0-\bar{\lambda}_1)\Gamma(\bar{\lambda}_1)C_{\mathcal{F},1}^{-\bar{\lambda}_1}}{\parallel\Lambda\parallel\prod_{i=1}^{n+1}\Gamma(\lambda_i+1)}\int \prod_{a=2}^n\mathrm{d}u_a\\
&\left(C_{\mathcal{F},n+1}+\sum_{a=2}^{n}C_{\mathcal{F},a}u_a\right)^{\lambda_0+\bar{\lambda}_1}\prod_{a=2}^{n}u_a^{\bar{\lambda}_a-1}\\
=&\dots\\
=&\frac{(L+1)\prod_{a=1}^{n+1}\left[\Gamma(\bar{\lambda}_a)C_{\mathcal{F},a}^{-\bar{\lambda}_a}\right]}{\parallel\Lambda\parallel\prod_{i=1}^{n+1}\Gamma(\lambda_i+1)}.
\end{split}
\end{equation}

\subsection{Analytic continuation}
While the analytic continuation is not a problem for the calculations in this paper, it needs to be considered in order to develop a general-purpose algorithm. For a $\cF$ polynomial with both positive terms and negative terms, a Feynman parameter may cross a branch point in the region of integration. Generally speaking, it is not easy to determine the branch while a Feynman parameter crosses a branch point. A possible solution to this problem is as follows. We replace each negative coefficient of $\cF$, denoted by $-C_{\cF,a}$, by $-yC_{\cF,a}$, and construct differential equations with respect to $y$. The imaginary part of $y$ should be $i0^+$ due to the $i0^+$ prescription of Feynman propagators. We determine the boundary conditions at $y=0^-$. All the boundary integrals are with positive definite $\cF$ polynomials and thus can further be evaluated by using the method described in previous subsections.

\subsection{Examples}
As an example of the application of the method described in this section, we consider the calculation of the following integral:
\begin{equation*}
\begin{split}
    &I_1(-\frac{d}{2},0,0,0,0,0,0,0)\\
    =&-\frac{i}{\pi^{3d/2}}\int\mathrm{d}^dl_1\mathrm{d}^dl_2\mathrm{d}^dl_3\frac{1}{l_1^+l_3^2\left(l_1-q\right)^2(q^--l_1^-)\left(l_2-q\right)^2 \left(l_1-l_3\right)^2 \left(l_2-l_3\right)^2}.
\end{split}
\end{equation*}
The $\mathcal{F}$ polynomial for this topology reads
\begin{equation*}
\begin{split}
\mathcal{F}_1=&x_8\left(x_{2,3,5}+x_{2,3,7}+x_{2,5,6}+x_{2,6,7}+x_{3,5,6}+x_{3,5,7}+x_{3,6,7}+x_{5,6,7}\right)\\
&-\left(x_{1,2,3,5}+x_{1,2,3,7}+x_{1,2,4,5}+x_{1,2,4,7}+x_{1,3,5,6}+x_{1,3,5,7}+x_{1,3,6,7}\right.\\
&\left.+x_{1,4,5,6}+x_{1,4,5,7}+x_{1,4,6,7}+x_{1,5,6,7}+x_{2,4,5,6}+x_{2,4,6,7}\right).
\end{split}
\end{equation*}
Here we use $x_{i,j,\dots}$ to denote $x_ix_j\cdots$. The $\bm{k}_r$ vectors for this polynomial can be found by using \texttt{Qhull}~\cite{Barber:1996,qhull}. Due to the homogeneity of the integrand of a parametric integral, two vectors $\bm{k}_{r_1}$ and $\bm{k}_{r_2}$ describe the same region if $k_{r_2,0}=k_{r_1,0}+(L+1)c$ and $k_{r_2,i}=k_{r_1,i}+c,~i\neq0$, for an arbitrary constant $c$. We fix this ambiguity with the constraint $k_{r,n+1}=0$. We get
\begin{equation*}
\begin{pmatrix}
 0 & 0 & 0 & 1 & 0 & 0 & 0 & 0 & 0 \\
 3 & 1 & 1 & 1 & 0 & 1 & 1 & 1 & 0 \\
 -1 & 0 & 0 & 0 & 0 & -1 & 0 & 0 & 0 \\
 -1 & 0 & 0 & 0 & 0 & 0 & -1 & 0 & 0 \\
 -3 & 0 & 0 & -1 & -1 & -1 & -1 & -1 & 0 \\
 -2 & 0 & -1 & 0 & 0 & -1 & 0 & -1 & 0 \\
 -1 & 0 & 0 & 0 & 0 & 0 & 0 & -1 & 0 \\
 3 & 0 & 1 & 1 & 1 & 1 & 1 & 1 & 0 \\
 -3 & 0 & -1 & -1 & 0 & -1 & -1 & -1 & 0 \\
 -2 & -1 & -1 & 0 & 0 & 0 & -1 & 0 & 0 \\
 -4 & -1 & -1 & -1 & -1 & -1 & -1 & -1 & 0 \\
 1 & 0 & 0 & 0 & 0 & 1 & 0 & 1 & 0 \\
 -1 & 0 & 0 & -1 & -1 & 0 & 0 & 0 & 0
\end{pmatrix}.
\end{equation*}
Here each row represents $(k_{r,0},k_{r,1},\dots,k_{r,n+1})$. There are $7$ pairs $\{i,~j\}$ with only one $\bm{k}_r$ such that $k_{r,i}>k_{r,j}$, which are $\{2,~1\}$, $\{4,~3\}$, $\{5,~7\}$, $\{6,~1\}$, $\{6,~2\}$, $\{6,~3\}$, and $\{7,~5\}$. Among these pairs, the pairs $\{5,~7\}$ and $\{7,~5\}$ has the minimal $N_r$. We choose the pair $\{7,~5\}$. The $\bm{k}_r$ vector with $k_{r,7}>k_{r,5}$ is $(-1,0,0,0,0,-1,0,0,0 )$.

After inserting $\delta(y-\frac{x_5}{x_7})$ and eliminating the $x_5$ integration, we get a $y$-dependent integral
\begin{equation*}
    I_{2,0}=I_2(-\frac{d}{2},0,0,0,0,0,1),
\end{equation*}
with the $\mathcal{F}$ polynomial
\begin{align*}
    \mathcal{F}_2=&x_7\left[y \left(x_{2,3,6}+x_{2,5,6}+x_{3,5,6}+x_3 x_6^2+x_5 x_6^2\right)+x_{2,3,6}+x_{2,5,6}+x_{3,5,6}\right]\\
    &-y\left(x_6^2 x_{1,3}+x_6^2 x_{1,4}+x_6^2 x_{1,5}+x_{1,2,3,6}+x_{1,2,4,6}+x_{1,3,5,6}+x_{1,4,5,6}+x_{2,4,5,6}\right)\\
    &-\left(x_{1,2,3,6}+x_{1,2,4,6}+x_{1,3,5,6}+x_{1,4,5,6}+x_{2,4,5,6}\right).
\end{align*}
By construction, we can easily get the momentum-space correspondence
\begin{equation*}
    I_{2,0}=-\frac{i}{\pi^{3d/2}}\int\mathrm{d}^dl_1\mathrm{d}^dl_2\mathrm{d}^dl_3\frac{1}{l_1^+l_3^2\left(l_1-q\right)^2(q^--l_1^-)\left(l_1-l_3\right)^2\left[y\left(l_2-q\right)^2+\left(l_2-l_3\right)^2\right]^2}.
\end{equation*}
This integral can be further reduced. We have
\begin{align*}
    I_{2,0}=-\frac{(3 d-8) (5 d-16) (5 d-14) (y+1)}{4 (d-3)^2 y} I_{2,1}+\frac{(d-2) (2 d-7) (3 d-8) (y+1)}{4 (d-3)^2 y} I_{2,2},
\end{align*}
where the master integrals are
\begin{align*}
I_{2,1}=&I_2(-\frac{d}{2},0,0,0,0,0,0),\\
I_{2,2}=&I_2(-\frac{d}{2},0,0,-1,0,0,0).
\end{align*}
The differential equations for these integrals are quite simple:
\begin{align*}
    \frac{\partial}{\partial y}I_{2,i}=\frac{y(\epsilon -2)-\epsilon +1}{y (y+1)}I_{2,i},\quad i=1,~2,
\end{align*}
where $\epsilon\equiv\frac{1}{2}(4-d)$. This differential equation can be trivially solved. The boundary conditions are determined by expanding the master integrals asymptotically in the limit of $y\to0$. We consider the integral $I_{2,1}$ for example. As is already known, there is only one region, for which the Feynman parameters scale as
\begin{align*}
    &x_6\sim y^{-1},\\
    &x_i\sim 1, \qquad i\neq 6.
\end{align*}
Rescaling the $\mathcal{F}$ polynomial according to the above scaling, and expanding it to the leading order in $y$, we get
\begin{equation*}
    \lim_{y\to0}I_{2,2}\to y^{1-\epsilon}I_{3}(-\frac{d}{2},0,0,0,0,0,0).
\end{equation*}
The $\mathcal{F}$ polynomial for the integral family $I_3$ is
\begin{align*}
    \mathcal{F}_3=&x_7\left(x_{2,3,6}+x_{2,5,6}+x_{3,5,6}+x_3 x_6^2+x_5 x_6^2\right)\\
    &-\left(x_6^2 x_{1,3}+x_6^2 x_{1,4}+x_6^2 x_{1,5}+x_{1,2,3,6}+x_{1,2,4,6}+x_{1,3,5,6}+x_{1,4,5,6}+x_{2,4,5,6}\right).
 \end{align*}
The integral family $I_3$ does not have an evident momentum-space correspondence. It can further be calculated by using the method described in this section. The $\mathbf{k}_r$ vectors for $\mathcal{F}_3$ are
\begin{align*}
\begin{pmatrix}
 0 & 0 & 0 & 1 & 0 & 0 & 0 & 0 \\
 -3 & 0 & 0 & -1 & -1 & -1 & -1 & 0 \\
 -4 & -1 & -1 & -1 & -1 & -1 & -1 & 0 \\
 -1 & 0 & 0 & 0 & 0 & -1 & 0 & 0 \\
 -2 & -1 & -1 & 0 & 0 & -1 & 0 & 0 \\
 -3 & 0 & -1 & -1 & 0 & -1 & -1 & 0 \\
 -2 & 0 & -1 & 0 & 0 & 0 & -1 & 0 \\
 3 & 0 & 1 & 1 & 1 & 1 & 1 & 0 \\
 3 & 1 & 1 & 1 & 0 & 1 & 1 & 0 \\
 1 & 0 & 0 & 0 & 0 & 0 & 1 & 0 \\
 -1 & 0 & 0 & -1 & -1 & 0 & 0 & 0
 \end{pmatrix}.
\end{align*}
There are $7$ pairs $\{x_i,~x_j\}$ with only one $\bm{k}_r$ such that $k_{r,i}>k_{r,j}$, among which the pair $\{2,~6\}$ has the minimal $N_r$. Insertion of $\delta(y-\frac{x_2}{x_6})$ leads to the integral
\begin{equation*}
    I_{4,0}=I_4(-\frac{d}{2},0,0,0,0,1),
\end{equation*}
with the $\mathcal{F}$ polynomial
\begin{align*}
    \mathcal{F}_4=&x_6\left[y \left(x_ 2 x_ 5^2+x_ 4 x_ 5^2\right)+x_{2,4,5}+x_2 x_5^2+x_4 x_5^2\right]-y \left(x_ 5^2 x_{1,2}+x_ 5^2 x_{1,3}+x_ 5^2 x_{3,4}\right)\\
    &-\left(x_5^2 x_{1,2}+x_5^2 x_{1,3}+x_5^2 x_{1,4}+x_{1,2,4,5}+x_{1,3,4,5}\right].
\end{align*}
The integral $I_{4,0}$ can further be reduced to
\begin{equation*}
    I_{4,0}=-\frac{9 (d-2) (y+1)^2}{4 y^2} I_{4,1}-\frac{3 (d-3)}{y} I_{4,2},
\end{equation*}
where the master integrals are
\begin{align*}
    I_{4,1}=&I_4\left(-\frac{d}{2},0,-1,0,0,0\right),\\
    I_{4,2}=&I_4\left(-\frac{d}{2},0,0,0,0,0\right).
\end{align*}
The differential equations for these integrals are
\begin{align*}
    \frac{\partial}{\partial y}
    \begin{pmatrix}
    I_{4,1}\\
    I_{4,2}
    \end{pmatrix}
    =
    \begin{pmatrix}
    \frac{(2 y-3) (\epsilon -1)}{y (y+1)} & 0 \\
    \frac{3 (y+1) (\epsilon -1)}{y^2} & -\frac{(y+2) (\epsilon -1)}{y \
(y+1)} 
    \end{pmatrix}
    .
    \begin{pmatrix}
    I_{4,1}\\
    I_{4,2}
    \end{pmatrix}.
\end{align*}
This differential-equation system can be converted into the canonical form~\cite{Henn:2013pwa} by using the package \texttt{epsilon}~\cite{Prausa:2017ltv} which implements Lee's algorithm~\cite{Lee:2014ioa}. The obtained differential-equation system is solved by using the standard differential-equation method. The boundary conditions can be determined by applying the method developed in this section recursively. Here we do not go into more detail.

\section{Soft theorem at three loops in QCD and ${\cal N}=4$ sYM}\label{sec:SoftTheorThrLoopsQCDSYM}

In this section, we present the results for the three-loop soft factor in QCD up to ${\cal O}(\epsilon^2)$. These results are necessary ingredients for QCD corrections or soft function calculation at N$^4$LO. We also derive the corresponding soft factor in ${\cal N} = 4$ sYM with full-color dependence, using the principle of leading transcendentality~\cite{Kotikov:2002ab}. Note that the principle of leading transcendentality has not been proved, but is known to work in many cases, including e.g. twist operator dimensions~\cite{Kotikov:2002ab,Kotikov:2004er}, soft functions or Wilson loops~\cite{Li:2014afw,Li:2016ctv}, form factors~\cite{Jin:2018fak,Ahmed:2019yjt,Lee:2022nhh}. We verify that the leading color contributions agree with a previous calculation~\cite{Li:2013lsa} based on BDS ansatz~\cite{Bern:2005iz}, and determine a three-loop constant $f_2^{(3)}$ analytically. 

\subsection{IR singularities of soft factor}
\label{subsec:IRSin}

Before presenting our results for soft factor at three loops, we first discuss its infrared singularities. IR singularities for scattering amplitudes have been understood to be factorized, as a result of soft-collinear factorization~\cite{Catani:1998bh,Becher:2009cu,Gardi:2009qi}. Since the soft factor is simply the soft limit of scattering amplitude, we can extract the IR singularities of the soft factor by taking the soft limit in the IR singularities of scattering amplitude. 

To all orders in perturbation theory, the IR singularities of massless scattering amplitudes are governed by a multiplicative renormalization factor $\boldsymbol{Z}$, which in general is a matrix in color space
\begin{equation}
\label{eq:Z}
| {\cal M}_{\text{fin.},n} (\{ p \}, \mu) \rangle = \lim_{\epsilon \to 0} \boldsymbol{Z}^{-1}(\epsilon, \{p\}, \mu) | {\cal M}_n (\epsilon, \{ p \}) \rangle,
\end{equation}
where in the equation above
$| {\cal M}_{\text{fin.},n} (\{ p \}, \mu) \rangle$ is the IR renormalized finite amplitude, and $| {\cal M}_n (\epsilon, \{ p \}) \rangle$ is the UV renormalized amplitude. The IR renormalized factor is
\begin{equation}
    \boldsymbol{Z}(\epsilon, \{p\}, \mu)
    = \mathbf{P} \exp \left[ \int_\mu^\infty \frac{d\mu'}{\mu'} \boldsymbol{\Gamma} (\{p \}, \mu')\right]
\end{equation}
and
\begin{equation}
\label{eq:GammaDefi}
    \boldsymbol{\Gamma}(\{p\},\mu) = \sum_{(i,j)} \frac{\boldsymbol{T}_i \cdot \boldsymbol{T}_j}{2} \gamma^{\rm cusp} (\alpha_s) \ln \frac{\mu^2}{-s_{ij}} + \sum_i \gamma^i(\alpha_s) + \boldsymbol{\Delta_3} + \mathcal{O}(\boldsymbol{\Delta}_4)\,,
\end{equation}
where explicit data for the anomalous dimension will be provided in the appendix. 
The Mandelstam variable is defined as $s_{ij} = 2 \sigma_{j} p_i p_j + i0$ with the sign factor
\begin{equation}
    \sigma_{ij} = \begin{cases}
    +1 \,, \quad p_i, p_j \text{ both incoming or outgoing}
    \\
    -1 \,, \quad \text{otherwise}
    \end{cases}
\end{equation}
The notation $(i,j)$ refers to unordered pairs of distinct parton indices. In eq.~\eqref{eq:GammaDefi}, $\boldsymbol{\Delta_3}$ refers to tripole contribution which is kinematics-independent and starts to contribute at the three-loop order~\cite{Almelid:2015jia}: 
\begin{align}
    \boldsymbol{\Delta_3}^{(3)} = -16 f_{abe} f_{cde} \left(\zeta_5 + 2 \zeta_2 \zeta_3\right) \sum_{i=1}^3 \sum_{\substack{1\leq j < k \leq 3 \\ j,k\neq i}} \left\{ \mathbf{T}_{i}^a, \mathbf{T}_{i}^d  \right\} \mathbf{T}_{j}^b \mathbf{T}_{k}^c \,.
\end{align}
Another contribution from the quadrupole term $\boldsymbol{\Delta}_4$ involving four partons is also present at the three-loop order, for example in the three-loop four-parton scattering amplitude in $\mathcal{N}=4$ and QCD~\cite{Henn:2016jdu,Caola:2021rqz,Caola:2021izf, Caola:2022dfa}. However, because we only deal with the soft factor of two hard partons (only scattering amplitude involving three colored partons is required), it is not needed in this work.

The soft factor computed in this work refers to the soft gluon limit of a three-parton amplitude in QCD. We write the IR renormalization formula as
\begin{equation}
\label{eq:Zsoft}
| {\cal M}_{\text{fin.},3} (p_1, p_2, p_3, \mu) \rangle = \lim_{\epsilon \to 0} \boldsymbol{Z}^{-1}(\epsilon, p_1, p_2, p_3, \mu) | {\cal M}_3 (\epsilon, p_1, p_2, p_3) \rangle,
\end{equation}
where $| {\cal M}_{3} (\epsilon, p_1, p_2, p_3) \rangle$ is the UV renormalized amplitudes with three massless QCD partons, for example $\gamma^* \to q(p_1) \bar q(p_2) g(p_3)$. In the soft gluon limit, the soft gluon factorization demands that the IR renormalized amplitude factorizes as 
\begin{equation}
    \lim_{p_3^0 \to 0} | {\cal M}_{\text{fin.},3}(p_1, p_2, p_3, \mu) \rangle = 
    J(p_3, \mu)  |{\cal M}_{\text{fin.},2}(p_1, p_2, \mu) \rangle \,, 
\end{equation}
where $J(p_3, \mu)$ is the IR renormalized soft factor, and $ |{\cal M}_{\text{fin.},2}(p_1, p_2, \mu) \rangle$ is the IR renormalized 2-parton amplitude,
\begin{align}
    J(p_3, \mu) = &\ \lim_{\epsilon \to 0} Z_s^{-1} (\epsilon, p_3, \mu) J(\epsilon, p_3) \,,
    \\
    |{\cal M}_{\text{fin.},2}(p_1, p_2, \mu) \rangle = &\ \lim_{\epsilon \to 0} Z_2^{-1} (\epsilon, p_1, p_2, \mu) |{\cal M}_2(\epsilon, p_1, p_2) \rangle \,.
\end{align}
This leads to the relation
\begin{equation}
    Z_s^{-1}(\epsilon, p_3, \mu) = \lim_{p_3 \to 0} Z_3^{-1}(\epsilon, p_1, p_2, p_3, \mu) Z_2(\epsilon, p_1, p_2, \mu)
\end{equation}
The infrared singularities of the three-loop soft factor can then be read-off from $Z_s(\epsilon, p_3, \mu)$.

\subsection{Soft theorem to three loops in QCD}
We are now ready to present our results for the soft factor to three loops. The results were verified to satisfy the (generalized) Casimir scaling principle, such that we are able to write them down in a unified form for both fundamental and adjoint representations. The corrections of $B_{12}$ in eq.~\eqref{eq:ColorDecomposition} were calculated to two loops in~\cite{Li:2013lsa,Duhr:2013msa}, we list them here using the convention of eq.~\eqref{eq:factorOutSe} for completeness. At one-loop order, the result can be expressed in terms of the following gamma functions, 
\begin{align}
\label{eq:b121}
    b_{12}^{(1)} = -\frac{\text{exp}\left(\gamma_{\text{E}}  \epsilon \right) \Gamma^3(1-\epsilon ) \Gamma^2(\epsilon +1)}{\epsilon ^2 \Gamma (1-2 \epsilon )}\,. 
\end{align}
For the two-loop corrections, we give the result to $\epsilon^4$ and found full agreement with the $\epsilon$ expansion of all-order result in~\cite{Duhr:2013msa},
\begin{align}
\label{eq:twoloopresults}
    b_{12}^{(2)}  = &\textcolor{blue}{C_A}  \bigg\{\frac{1}{2 \epsilon ^4} -\frac{11}{12 \epsilon ^3}+\frac{\zeta
   _2-\frac{67}{36}}{\epsilon ^2}+\frac{-\frac{11 \zeta _2}{12}-\frac{11 \zeta
   _3}{6}-\frac{193}{54}}{\epsilon }-\frac{67 \zeta _2}{36}+\frac{341 \zeta
   _3}{18}+\frac{7 \zeta _4}{8}-\frac{571}{81} \nonumber 
   \\
   &+ \epsilon \Big[-\frac{7}{6} \zeta _3 \zeta
   _2-\frac{139 \zeta _2}{54}+\frac{2077 \zeta _3}{54}+\frac{2035 \zeta
   _4}{48}-\frac{247 \zeta _5}{10}-\frac{3410}{243}\Big] + \epsilon^2 \Big[ -\frac{205
   \zeta _3^2}{18}  \nonumber 
   \\
   & +\frac{341 \zeta _2 \zeta _3}{18} +\frac{6388 \zeta
   _3}{81}-\frac{436 \zeta _2}{81}+\frac{12395 \zeta _4}{144}+\frac{5621 \zeta
   _5}{30}-\frac{3307 \zeta _6}{48}-\frac{20428}{729}\Big]  \nonumber 
   \\
   & + \epsilon^3 \Big[ -\frac{10571 \zeta _3^2}{54}+\frac{2077 \zeta _2 \zeta _3}{54}-\frac{509
   \zeta _4 \zeta _3}{24}+\frac{37427 \zeta _3}{243}-\frac{2411 \zeta
   _2}{243}+\frac{41105 \zeta _4}{216}  \nonumber 
   \\
   & -\frac{219 \zeta _2 \zeta _5}{10}+\frac{34237
   \zeta _5}{90}+\frac{42361 \zeta _6}{64}-\frac{4573 \zeta
   _7}{14}-\frac{122504}{2187}\Big] + \epsilon ^4 \Big[ -40 \zeta
   _{5,3}-\frac{845}{18} \zeta _2 \zeta _3^2 \nonumber 
   \\
   &-\frac{64387 \zeta _3^2}{162}+\frac{5524
   \zeta _2 \zeta _3}{81}-\frac{63085 \zeta _4 \zeta _3}{72}-\frac{29 \zeta _5 \zeta
   _3}{15}+\frac{226405 \zeta _3}{729}-\frac{14785 \zeta _2}{729}+\frac{119135 \zeta
   _4}{324} \nonumber 
   \\
   &+\frac{5621 \zeta _2 \zeta _5}{30}+\frac{108748 \zeta
   _5}{135}+\frac{258017 \zeta _6}{192}+\frac{90101 \zeta _7}{42}-\frac{1264777 \zeta
   _8}{1152}-\frac{734896}{6561}\Big] \bigg\}  \nonumber 
   \\
   & + \textcolor{blue}{N_f} \bigg\{\frac{1}{6 \epsilon ^3}+ \frac{5}{18 \epsilon ^2}+ \frac{\frac{\zeta
   _2}{6}+\frac{19}{54}}{\epsilon } + \frac{5 \zeta _2}{18}-\frac{31 \zeta
   _3}{9}+\frac{65}{162}+ \epsilon \Big[-\frac{35 \zeta _2}{54}-\frac{155 \zeta
   _3}{27} \nonumber 
   \\
   & -\frac{185 \zeta _4}{24}+\frac{211}{486}\Big]
   + \epsilon^2 \Big[-\frac{31}{9} \zeta _3 \zeta _2-\frac{367 \zeta _2}{162}-\frac{994 \zeta
   _3}{81}-\frac{925 \zeta _4}{72}-\frac{511 \zeta _5}{15}+\frac{665}{1458}\Big]  \nonumber 
   \\
   &+ \epsilon^3 \Big[\frac{961 \zeta _3^2}{27}-\frac{155 \zeta _2 \zeta
   _3}{27}-\frac{5255 \zeta _3}{243}-\frac{3083 \zeta _2}{486}-\frac{8915 \zeta
   _4}{216}-\frac{511 \zeta _5}{9}-\frac{3851 \zeta _6}{32}+\frac{2059}{4374}\Big] \nonumber 
   \\
   & + \epsilon ^4 \Big[\frac{4805 \zeta _3^2}{81}-\frac{130 \zeta _2 \zeta
   _3}{81}+\frac{5735 \zeta _4 \zeta _3}{36}-\frac{31246 \zeta _3}{729}-\frac{20503
   \zeta _2}{1458}-\frac{55225 \zeta _4}{648}-\frac{511 \zeta _2 \zeta
   _5}{15} \nonumber 
   \\
   &-\frac{19834 \zeta _5}{135}-\frac{19255 \zeta _6}{96}-\frac{8191 \zeta
   _7}{21}+\frac{6305}{13122}\Big]  \bigg\}\,.
\end{align}
The three-loop corrections to eq.~\eqref{eq:ColorDecomposition} are our main results, we expand the results to $\epsilon^2$ below: 
\begin{align}
    b_{12}^{(3)} = &\textcolor{blue}{C_A^2} \bigg\{-\frac{1}{6 \epsilon ^6}+ \frac{11}{12 \epsilon
   ^5}+ \frac{\frac{119}{324}-\frac{3 \zeta _2}{4}}{\epsilon ^4}+ \frac{\frac{649 \zeta
   _2}{216}+\frac{2 \zeta _3}{3}-\frac{1517}{486}}{\epsilon ^3} \nonumber 
   \\
   &+\frac{\frac{2501
   \zeta _2}{648}-\frac{2101 \zeta _3}{108}-\frac{1487 \zeta
   _4}{288}-\frac{7271}{486}}{\epsilon ^2}+ \frac{\frac{11 \zeta _3 \zeta
   _2}{18}+\frac{437 \zeta _2}{972}+\frac{2575 \zeta _3}{36}-\frac{22583 \zeta
   _4}{576}+\frac{98 \zeta _5}{5}-\frac{446705}{8748}}{\epsilon }\nonumber 
   \\
   &+\frac{293 \zeta
   _3^2}{36}-\frac{2453 \zeta _2 \zeta _3}{72}+\frac{203705 \zeta
   _3}{486}-\frac{12911 \zeta _2}{2916}+\frac{493381 \zeta _4}{1728}-\frac{26543
   \zeta _5}{60}+\frac{445679 \zeta
   _6}{6912}\nonumber 
   \\
   &-\frac{8206861}{52488}+\epsilon \Big[-\frac{17149 \zeta _3^2}{216}+\frac{21031
   \zeta _2 \zeta _3}{216}+\frac{86 \zeta _4 \zeta _3}{9}+\frac{2330483 \zeta
   _3}{1458}-\frac{403379 \zeta _2}{17496}\nonumber 
   \\
   &+\frac{1228523 \zeta _4}{864}+\frac{9773
   \zeta _2 \zeta _5}{90}+\frac{262597 \zeta _5}{180}-\frac{25965643 \zeta
   _6}{13824}+\frac{151631 \zeta _7}{252}-\frac{48027739}{104976}\Big] \nonumber 
   \\
   &+ \epsilon ^2 \Big[-\frac{15008 \zeta _{5,3}}{45}+\frac{10045}{72} \zeta _2 \zeta
   _3^2-\frac{920995 \zeta _3^2}{216}+\frac{71831 \zeta _2 \zeta
   _3}{108}+\frac{388289 \zeta _4 \zeta _3}{576}-\frac{9907 \zeta _5 \zeta
   _3}{30}\nonumber 
   \\
   &+\frac{15854467 \zeta _3}{2916}-\frac{5363867 \zeta
   _2}{104976}+\frac{42678481 \zeta _4}{7776}-\frac{71533 \zeta _2 \zeta
   _5}{120}+\frac{82837 \zeta _5}{10}+\frac{112195243 \zeta _6}{13824}\nonumber 
   \\
   &-\frac{1343045
   \zeta _7}{126}+\frac{3738034847 \zeta
   _8}{829440}-\frac{2482106477}{1889568}\Big]\bigg\} 
   + \textcolor{blue}{C_A N_f} \bigg\{-\frac{1}{6 \epsilon ^5}+ \frac{43}{162 \epsilon
   ^4}\nonumber 
   \\
   &+ \frac{\frac{895}{486}-\frac{59 \zeta _2}{108}}{\epsilon ^3}+\frac{-\frac{31
   \zeta _2}{324}+\frac{239 \zeta _3}{54}+\frac{2603}{486}}{\epsilon
   ^2}+\frac{\frac{3265 \zeta _2}{972}-\frac{4945 \zeta _3}{162}+\frac{2437 \zeta
   _4}{288}+\frac{24169}{2187}}{\epsilon }+ \frac{271 \zeta _3 \zeta
   _2}{36}\nonumber 
   \\
   &-\frac{3925 \zeta _2}{2916}-\frac{2513 \zeta _3}{18}-\frac{33109 \zeta
   _4}{288}+\frac{7799 \zeta _5}{90}+\frac{397699}{26244}+ \epsilon \Big[-\frac{4969 \zeta_3^2}{108}-\frac{1595 \zeta _2 \zeta _3}{36}\nonumber 
   \\
   & -\frac{720299 \zeta
   _3}{1458}-\frac{228895 \zeta _2}{4374}-\frac{1168171 \zeta _4}{2592}-\frac{187753
   \zeta _5}{270}+\frac{2476865 \zeta _6}{6912}-\frac{22273}{5832}\Big]
   \nonumber 
   \\
   &+ \epsilon ^2 \Big[\frac{404075 \zeta _3^2}{324}-\frac{78295 \zeta _2 \zeta
   _3}{324}-\frac{121555 \zeta _4 \zeta _3}{288}-\frac{3316207 \zeta
   _3}{2187}-\frac{17477627 \zeta _2}{52488}\nonumber 
   \\
   &-\frac{15232813 \zeta
   _4}{7776}+\frac{7063 \zeta _2 \zeta _5}{60}-\frac{52115 \zeta
   _5}{18}-\frac{76597939 \zeta _6}{20736}+\frac{13871 \zeta
   _7}{7}-\frac{125652667}{944784}\Big] \bigg\} \nonumber 
   \\
   &
   +\textcolor{blue}{ C_F N_f} \bigg\{\frac{1}{9 \epsilon
   ^3}+\frac{\frac{55}{54}-\frac{8 \zeta _3}{9}}{\epsilon
   ^2}+\frac{\frac{\zeta _2}{6}-\frac{76 \zeta
   _3}{27}-\frac{4 \zeta _4}{3}+\frac{1819}{324}}{\epsilon
   }-\frac{4}{3} \zeta _3 \zeta _2+\frac{67 \zeta
   _2}{36}-\frac{1385 \zeta _3}{81}\nonumber 
   \\
   &-\frac{38 \zeta
   _4}{9}-\frac{56 \zeta
   _5}{9}+\frac{45967}{1944}+\epsilon \Big[\frac{544 \zeta
   _3^2}{9}-\frac{38 \zeta _2 \zeta _3}{9}-\frac{50495 \zeta
   _3}{486}+\frac{3547 \zeta _2}{216}-\frac{16237 \zeta
   _4}{432} \nonumber 
   \\
   & -\frac{532 \zeta _5}{27}-\frac{101 \zeta
   _6}{6}+\frac{1007179}{11664}\Big]+
   \epsilon^2 \Big[\frac{5168 \zeta _3^2}{27}-\frac{809 \zeta _2 \zeta
   _3}{54}+\frac{599 \zeta _4 \zeta _3}{2}-\frac{1661303
   \zeta _3}{2916}\nonumber 
   \\
   &+\frac{99931 \zeta _2}{1296}-\frac{635899
   \zeta _4}{2592}-\frac{28 \zeta _2 \zeta _5}{3}-\frac{70417
   \zeta _5}{405}-\frac{1919 \zeta _6}{36}-\frac{392 \zeta
   _7}{9}+\frac{20357263}{69984}\Big] \bigg\} \nonumber 
   \\
   & 
   + \textcolor{blue}{N_f^2} \bigg\{-\frac{4}{81 \epsilon ^4}+ -\frac{40}{243 \epsilon
   ^3}+ \frac{-\frac{2 \zeta _2}{27}-\frac{8}{27}}{\epsilon
   ^2}+ \frac{-\frac{20 \zeta _2}{81}+\frac{260 \zeta
   _3}{81}-\frac{704}{2187}}{\epsilon }+ \frac{44 \zeta
   _2}{27}+\frac{2600 \zeta _3}{243}\nonumber 
   \\
   &+\frac{1229 \zeta
   _4}{108}+\frac{640}{6561}+ \epsilon \Big[\frac{130 \zeta _3 \zeta
   _2}{27}+\frac{5984 \zeta _2}{729}+\frac{296 \zeta
   _3}{9}+\frac{6145 \zeta _4}{162}+\frac{10084 \zeta
   _5}{135}+\frac{12160}{6561}\Big]\nonumber 
   \\
   &
   +\epsilon^2 \Big[-\frac{8450 \zeta _3^2}{81}+\frac{1300 \zeta _2
   \zeta _3}{81}+\frac{168448 \zeta _3}{2187}+\frac{67712
   \zeta _2}{2187}+\frac{9355 \zeta _4}{54}+\frac{20168 \zeta
   _5}{81}\nonumber 
   \\
   &+\frac{999593 \zeta
   _6}{2592}+\frac{423296}{59049}\Big] \bigg\}\,,
\end{align}

\begin{align}
    c_{12}^{(3)} =& \frac{-32 \zeta _2 \zeta _3-16 \zeta
   _5}{\epsilon }  -192 \zeta _3^2+\frac{64 \zeta _3}{3}-64
   \zeta _2\nonumber 
   \\
   &+\frac{1760 \zeta _5}{3}-940 \zeta
   _6+ \epsilon\Big[\frac{4928 \zeta _3^2}{3}-1112 \zeta _4 \zeta
   _3-\frac{1696 \zeta _3}{9}-416 \zeta _2+208 \zeta _4-1496
   \zeta _2 \zeta _5\nonumber 
   \\
   &+\frac{10720 \zeta _5}{9}+1760 \zeta
   _6-4032 \zeta _7\Big] + \epsilon ^2
   \Big[\frac{29376 \zeta _{5,3}}{5}-480 \zeta _2 \zeta
   _3^2+\frac{30016 \zeta _3^2}{9}+608 \zeta _2 \zeta _3\nonumber 
   \\
   &+4928
   \zeta _4 \zeta _3+5488 \zeta _5 \zeta _3-\frac{42560 \zeta
   _3}{27}-\frac{6208 \zeta _2}{3}-\frac{10048 \zeta
   _4}{3}+880 \zeta _2 \zeta _5+\frac{101216 \zeta
   _5}{27}\nonumber 
   \\
   &+\frac{10720 \zeta _6}{3}+27280 \zeta
   _7-\frac{2613298 \zeta _8}{45}\Big] \,, 
\end{align}

\begin{align}
\label{eq:d123}
    d_{12}^{(3)} = &128 \zeta _2-\frac{128 \zeta
   _3}{3}-\frac{640 \zeta _5}{3}\nonumber 
   \\
   &+\epsilon \Big[-\frac{1792 \zeta
   _3^2}{3}+\frac{3008 \zeta _3}{9}+960 \zeta _2-416 \zeta
   _4-\frac{3200 \zeta _5}{9}-640 \zeta _6\Big] 
   + \epsilon^2 \Big[-\frac{8960 \zeta _3^2}{9}\nonumber 
   \\
   &-1216 \zeta _2 \zeta
   _3-1792 \zeta _4 \zeta _3+\frac{94144 \zeta
   _3}{27}+\frac{15296 \zeta _2}{3}+\frac{18848 \zeta
   _4}{3}-320 \zeta _2 \zeta _5\nonumber 
   \\
   &-\frac{91072 \zeta
   _5}{27}-\frac{3200 \zeta _6}{3}-9920 \zeta _7\Big]\,,
\end{align}
where several regular zeta values up to transcendental-weight 8 and one multiple zeta value are involved, 
\begin{align}
    \zeta_{5,3} = \sum_{m=1}^\infty \sum_{n=1}^{m-1} \frac{1}{m^5 n^3} \simeq 0.0377076729848475 \,.
\end{align}
The higher-order corrections of the Eikonal functions in eq.~\eqref{eq:eikonalfunction} can be readily expressed in terms of the above results, in the cases of one and two loops,
\begin{align}
    r_{12}^{(l)} = C_A b_{12}^{(l)} \, \text{ for } l=1,\, 2 \,, 
\end{align}
and in the three-loop case,
\begin{align}
\label{eq:threeLoopR12}
    r_{12}^{(3)} = C_A \, b_{12}^{(3)} + \frac{d_R^{abcd} d_A^{abcd}}{N_R C_R}\, c_{12}^{(3)} + \frac{d_R^{abcd} d_F^{abcd} N_f}{N_R C_R} \, d_{12}^{(3)} \,,
\end{align}
where in gauge group SU($N_c$), the quadratic color structures are evaluated to be the following explicit expressions, 
\begin{align}
    &\frac{d_F^{abcd} d_F^{abcd}}{N_F C_F} =  \frac{N_c^4 - 6 N_c^2 +18}{48 N_c^2},\quad \frac{d_F^{abcd} d_A^{abcd}}{N_F C_F} =  \frac{N_c^3+ 6 N_c}{24}, \nonumber 
    \\
    & \frac{d_A^{abcd} d_F^{abcd}}{N_A C_A} = \frac{N_c^2+6}{48},\quad \frac{d_A^{abcd} d_A^{abcd}}{N_A C_A} =  \frac{N_c^3+ 36 N_c}{24}\,.
\end{align}
We emphasize that our results are for unrenormalized quantities. To perform the ultraviolet (UV) renormalization, we just need to renormalize the strong coupling constant, i.e., 
\begin{align}
    a_s \to  Z_{a_s} a_s\,,
\end{align}
with $Z_{a_s} = 1- \frac{\beta_0}{\epsilon}  a_s  + \left( \frac{\beta_0^2}{\epsilon^2} - \frac{\beta_1}{2 \epsilon}\right) + \left(   -\frac{\beta_0^3}{\epsilon^3} + \frac{7 \beta_1 \beta_0}{6 \epsilon^2} -\frac{\beta_2}{3 \epsilon}\right)+ \mathcal{O}(a_s^4)$, where $\beta_i$ is QCD beta function which will be collected in the appendix. After UV renormalization, the remaining poles stem from IR singularities. We checked that the IR poles in our explicit results agree with those predicted in section~\ref{subsec:IRSin}.

\subsection{Soft theorem in ${\cal N}=4$ sYM and BDS ansatz at three loops}
The soft theorem in ${\cal N}=4$ sYM can be easily extracted from QCD results assuming the principle of maximal transcendentality~\cite{Kotikov:2002ab,Kotikov:2004er}. At the one-loop order, the Eikonal function for the soft theorem in  ${\cal N}=4$ sYM and QCD are identical, i.e., 
\begin{align}
    S^{(1)}_{12,\,{\cal N} = 4}(q) = S_{12}^{(1)}(q)=  S_{12}^{0}(q) S_\epsilon \textcolor{blue}{N_c} \, b_{12}^{(1)}\,.
\end{align}

At two-loop order, by keeping only the maximal transcendentality part of eq.~\eqref{eq:twoloopresults}, the $\mathcal{N} =4$ sYM results up to $\epsilon^4$ reads,
\begin{align}
\label{eq:n4twoloopS}
    S^{(2)}_{12,\,{\cal N} = 4}(q) = \,& S_{12}^{0}(q) S_\epsilon^2 \textcolor{blue}{N_c^2} \bigg\{ \frac{1}{2 \epsilon ^4}+ \frac{\zeta _2}{\epsilon ^2} -\frac{11 \zeta _3}{6
   \epsilon }+ \frac{7 \zeta _4}{8}+ \epsilon \left(-\frac{7}{6} \zeta _2 \zeta _3-\frac{247
   \zeta _5}{10}\right)  \nonumber 
    \\
    & + \epsilon^2 \left(-\frac{205 \zeta _3^2}{18}-\frac{3307 \zeta
   _6}{48}\right) + \epsilon ^3 \left(-\frac{509}{24} \zeta _3 \zeta _4-\frac{219 \zeta
   _2 \zeta _5}{10}-\frac{4573 \zeta _7}{14}\right) \nonumber 
    \\
    & + \epsilon ^4 \left(-40
   \zeta _{5,3}-\frac{845}{18} \zeta _2 \zeta _3^2-\frac{29 \zeta _5 \zeta
   _3}{15}-\frac{1264777 \zeta _8}{1152}\right) \bigg\} \,,
\end{align}
where only the leading color contributes. Similarly, at the three-loop order, we have
\begin{align}
\label{eq:n4threeloopS}
    S^{(3)}_{12,\,{\cal N} = 4}(q) =& \, S_{12}^{0}(q) S_\epsilon^3 \Bigg[\textcolor{blue}{ N_c^3} \bigg\{-\frac{1}{6 \epsilon ^6}-\frac{3 \zeta _2}{4
   \epsilon ^4}+\frac{2 \zeta _3}{3 \epsilon ^3}-\frac{1487
   \zeta _4}{288 \epsilon ^2}+\frac{\frac{284 \zeta
   _5}{15}-\frac{13 \zeta _2 \zeta _3}{18}}{\epsilon
   }+ \frac{5 \zeta _3^2}{36}\nonumber 
    \\
    & +\frac{174959 \zeta
   _6}{6912}  +  \epsilon \left(-\frac{331}{9} \zeta _3 \zeta
   _4+\frac{4163 \zeta _2 \zeta _5}{90}+\frac{109295 \zeta
   _7}{252}\right)\nonumber 
    \\
    & + \epsilon ^2 \left(-\frac{3992
   \zeta _{5,3}}{45}+\frac{8605}{72} \zeta _2 \zeta
   _3^2-\frac{3047 \zeta _5 \zeta _3}{30}+\frac{1731021983
   \zeta _8}{829440}\right)\bigg\} \nonumber 
    \\
    & + \frac{3}{2} \textcolor{blue}{N_c} \bigg\{\frac{-32 \zeta _2 \zeta _3-16 \zeta
   _5}{\epsilon } -192 \zeta _3^2-940 \zeta _6+\epsilon \big(-1112
   \zeta _3 \zeta _4-1496 \zeta _2 \zeta _5 \nonumber
   \\
   &-4032 \zeta
   _7\big)   + \epsilon ^2 \left(\frac{29376 \zeta
   _{5,3}}{5}-480 \zeta _2 \zeta _3^2+5488 \zeta _5 \zeta
   _3-\frac{2613298 \zeta _8}{45}\right)\bigg\} \Bigg]\,,
\end{align}
where the three-loop $\mathcal{N} =4$ result receives contributions from both leading color and sub-leading color. The sub-leading color appears for the first time at three-loop order and comes from the fourth invariant tensor $d_A^{abcd} d_A^{abcd}$ in~\eqref{eq:threeLoopR12} solely. While not rigorously proved, the validity of the principle of maximal transcendentality at subleading color was confirmed in the case of the four-loop Sudakov form factor~\cite{Lee:2021lkc,Lee:2022nhh}.~\footnote{We note that the subleading-color soft current in $\mathcal{N}=4$ SYM was very recently obtained in \cite{Herzog:2023sgb} using an alternative approach, which involves performing an expansion-by-regions analysis starting from the full $\mathcal{N}=4$ form factors. Their results confirm that our predictions, which are based on the maximal transcendental principle, are correct.} 
Interestingly, the soft limit of three-loop splitting amplitude in planar $\mathcal{N} =4$ sYMs can be reorganized into the following form~\cite{Li:2013lsa} which is the soft limit of the well-known BDS ansatz~\cite{Bern:2005iz}, 
\begin{align}
    r_S^{(3)}(\epsilon) = -\frac{1}{3} \left( r_S^{(1)} (\epsilon)\right)^3 + r_S^{(1)}(\epsilon) r_S^{(2)}(\epsilon) + f^{(3)}(\epsilon) r_S^{(1)} (3 \epsilon) + \mathcal{O}(\epsilon)\,,
\end{align}
where $r_S^{(i)}$ are related to the Eikonal functions in the following way,
\begin{align}
   S^{(1)}_{12,\,{\cal N} = 4}(q) &= 2 S_{12}^{0}(q) S_\epsilon N_c\, r_S^{(1)}(\epsilon) \,, 
    S^{(2)}_{12,\,{\cal N} = 4}(q) = 4 S_{12}^{0}(q) S_\epsilon^2 N_c^2\, r_S^{(2)}(\epsilon)\,, \nonumber
   \\
   S^{(3)}_{12,\,{\cal N} = 4}(q) &= 8  S_{12}^{0}(q) S_\epsilon^3 N_c^3 r_S^{(3)} (\epsilon) + \text{sub-leading color contribution}\,,
\end{align}
and the $f^{(3)}(\epsilon)$ has been calculated to order $\epsilon^2$~\cite{Spradlin:2008uu}
\begin{align}
    f^{(3)}(\epsilon) = \frac{11 \zeta_4}{2} + (5 \zeta_2 \zeta_3 + 6 \zeta_5)\epsilon + f_2^{(3)} \epsilon^2 + \mathcal{O}(\epsilon^3)\,,
\end{align}
where $f_2^{(3)}$ is known numerically only with $ f_2^{(3)}=85.263\pm 0.004 $. By comparing the predicted result from BDS ansatz with our explicit result shown in~\eqref{eq:n4threeloopS}, we managed to determine the  $f_2^{(3)}$ analytically to be
\begin{align}
     f_2^{(3)}= 31 \zeta _3^2+\frac{1909 \zeta _6}{48} \simeq 85.25374611\,,
\end{align}
which agrees well with the numerical calculation of $f_2^{(3)}$ in~\cite{Spradlin:2008uu}.  

\section{Conclusion}

High energy scattering amplitudes in QCD with a gluon radiation admit a universal factorization formula in the soft gluon limit in terms of a soft factor and an amplitude with the soft gluon removed, commonly known as soft theorem. In this paper we present a calculation for the soft factor through three loops in the expansion of the strong coupling constant. Our calculation was carried out with restriction to only two hard partons in the scattering processes, which is relevant for several important processes in collider physics, such as Drell-Yan production, $e^+e^-$ annihilation to dijet, and 1+1 jet production in DIS. We present analytic results for the soft factor through to ${\cal O}(\epsilon^2)$ in dimensional regularization parameter, which is needed for infrared subtraction and soft function calculation to N$^4$LO in perturbation theory. 

The calculation was done by expressing the soft factor as a single-gluon matrix element of soft Wilson line operator in SCET, and constructing the corresponding integrand using modern Feynman integral techniques. The three-loop soft factor can be reduced to the calculation of 49 single-scale soft master integrals. We developed a systematic iterative approach based on Feynman parameter representation, differential equation on Feynman parameter, and IBP reduction in Feynman parameter representation to tackle these master integrals. We expect that our approach can be applied well to other single-scale master integrals for both loop and phase space integrals. We verify that the infrared poles of the three-loop soft factor agree with the general infrared factorization formula of QCD, providing a strong check to our calculation. 

As an application of our results, we obtain the soft factor in ${\cal N}=4$ sYM by assuming the principle of maximal transcendentality, which states that for certain quantities in QCD and ${\cal N}=4$ sYM, their leading transcendental part is the same in perturbation theory. The three-loop soft factor at leading color approximation in ${\cal N}=4$ sYM agrees with previously known results obtained from BDS ansatz. In addition, we also analytically determine a three-loop constant $f_2^{(3)} = 31\zeta_3^2 + 1909 \zeta_6/48$ in BDS ansatz, which was only known numerically. Our new results are the full-color dependence of the three-loop soft factor, which can be used to check the three-loop form factor $1\to 3$ decay once the relevant master integrals are known. Towards the full-color three-loop $1 \to 3$ form factor in QCD, we notice that the corresponding leading-color result became available quite recently~\cite{Gehrmann:2023jyv}.

\acknowledgments
We would like to thank Lance J. Dixon for useful discussions. W.C. and H.X.Z. were supported by National Natural Science Foundation of China under contract No. 11975200.
M.X.L. was supported by National Natural Science Foundation of China under contract No. U2230402. T.Z.Y. would like to acknowledge the European Research Council (ERC) for funding this work under the European Union's Horizon 2020 research and innovation programme grant agreement 101019620 (ERC Advanced Grant TOPUP). H.X.Z. would also like to express gratitude to the Erwin-Schrödinger Institute for Mathematical Physics for their hospitality during the program "Quantum Field Theory at the Frontiers of the Strong Interaction", where part of this work was completed.

\appendix
\section{Integral families}
\label{sec:integralsFamilies}
We define the following integrals, 
\begin{align}
    J(;\nu_1, \nu_2,\cdots,\nu_{15}) = (\mu^2 e^{\gamma_E})^{3 \epsilon}  \int \frac{d^d l_1 d^d l_2 d^d l_3}{i^3 \pi^{3 d/2}} \frac{1}{D_1^{\nu_1} D_2^{\nu_2}  \cdots D_{15}^{\nu_{15}} }, 
\end{align}
with the denominator sets taken from the following six integral families, 
\begin{align*}
\centering
\begin{array}{|c|c|c|c|c|c|c|}
\hline
  & 1 & 2 & 4 & 7 & 15 & 16 \\
  \hline
 D_1 & n_1\cdot l_1 & n_1\cdot l_3 & n_1\cdot l_3 & n_1\cdot l_1 & n_1\cdot l_2 & n_1\cdot l_2 \\
 D_2 & n_1\cdot l_3 & n_1\cdot \left(l_3-l_2\right) & n_1\cdot \left(l_3-l_2\right) &
   n_1\cdot \left(l_1-l_3\right) & n_1\cdot \left(q-l_1\right) & n_1\cdot
   \left(l_3-q\right) \\
 D_3 & n_1\cdot l_2 & n_1\cdot \left(l_3-l_1\right) & n_1\cdot \left(l_3-l_1\right) &
   n_1\cdot \left(l_2-l_3\right) & n_1\cdot \left(l_3-l_1\right) & n_1\cdot
   \left(l_3-l_1\right) \\
 D_4 & n_2\cdot \left(q-l_1\right) & n_2\cdot \left(q-l_3\right) &
   n_2\cdot \left(l_2-l_3\right) & n_2\cdot \left(q-l_1\right) &
   n_2\cdot \left(q-l_2\right) & n_2\cdot \left(q-l_2\right) \\
 D_5 & n_2\cdot \left(q-l_3\right) & n_2\cdot \left(q-l_2\right) &
   n_2\cdot \left(l_1-l_3\right) & n_2\cdot \left(q-l_3\right) &
   n_2\cdot l_1 & n_2\cdot l_1 \\
 D_6 & n_2\cdot \left(q-l_2\right) & n_2\cdot \left(q-l_1\right) &
   n_2\cdot \left(q-l_3\right) & n_2\cdot \left(q-l_2\right) &
   n_2\cdot \left(l_3-l_2\right) & n_2\cdot \left(l_3-l_2\right) \\
 D_7 & l_1^2 & l_1^2 & l_1^2 & l_1^2 & l_1^2 & l_1^2 \\
 D_8 & \left(l_1-q\right){}^2 & \left(l_1-q\right){}^2 & \left(l_1-q\right){}^2 &
   \left(l_1-q\right){}^2 & \left(l_1-q\right){}^2 & \left(l_1-q\right){}^2 \\
 D_9 & l_2^2 & l_2^2 & l_2^2 & l_2^2 & l_2^2 & l_2^2 \\
 D_{10} & \left(l_2-q\right){}^2 & \left(l_2-q\right){}^2 & \left(l_2-q\right){}^2 &
   \left(l_2-q\right){}^2 & \left(l_2-q\right){}^2 & \left(l_2-q\right){}^2 \\
 D_{11} & l_3^2 & l_3^2 & l_3^2 & l_3^2 & l_3^2 & l_3^2 \\
 D_{12} & \left(l_3-q\right){}^2 & \left(l_3-q\right){}^2 & \left(l_3-q\right){}^2 &
   \left(l_3-q\right){}^2 & \left(l_3-q\right){}^2 & \left(l_3-q\right){}^2 \\
 D_{13} & \left(l_1-l_2\right){}^2 & \left(l_1-l_2\right){}^2 &
   \left(l_1-l_2\right){}^2 & \left(l_1-l_2\right){}^2 & \left(l_1-l_3\right){}^2 &
   \left(l_1-l_3\right){}^2 \\
 D_{14} & \left(l_1-l_3\right){}^2 & \left(l_1-l_3\right){}^2 &
   \left(l_1-l_3\right){}^2 & \left(l_1-l_3\right){}^2 & \left(l_2-l_3\right){}^2 &
   \left(l_2-l_3\right){}^2 \\
 D_{15} & \left(l_2-l_3\right){}^2 & \left(l_2-l_3\right){}^2 &
   \left(l_2-l_3\right){}^2 & \left(l_2-l_3\right){}^2 & \left(l_1+l_2-l_3\right){}^2
   & \left(l_1+l_2-l_3\right){}^2 \\
   \hline
\end{array}\,,
\end{align*}
where all propagators have Feynman prescription $+ i 0^+$, for example $n_1 \cdot l_1 + i 0^+$.

\section{QCD beta function and anomalous dimensions}
To predict the infrared singularities as shown in sec.~\ref{subsec:IRSin}, we need the relevant anomalous dimensions and QCD beta function, which will be listed below for readers' convenience. 

The QCD beta function is defined as
\begin{equation}
\frac{d\alpha_s}{d\ln\mu} = \beta(\alpha_s) = -2\alpha_s \sum_{n=0}^\infty \left( \frac{\alpha_s}{4 \pi} \right)^{n+1} \, \beta_n \, ,
\end{equation}
and here we need to three-loop order~\cite{Tarasov:1980au}
\begin{align}
\beta_0 &= \frac{11}{3} C_A - \frac{4}{3} T_F N_f \, , \nn
\\
\beta_1 &= \frac{34}{3} C_A^2 - \frac{20}{3} C_A T_F N_f - 4 C_F T_F N_f \, ,\nn
\\
\beta_2 &= \left(\frac{158 C_A}{27}+\frac{44 C_F}{9}\right) N_f^2 T_F^2 +\left(-\frac{205 C_A
   C_F}{9}-\frac{1415 C_A^2}{27}+2 C_F^2\right) N_f T_F  +\frac{2857 C_A^3}{54}\,.
\end{align}

We perform a perturbative expansion for the anomalous dimension $\gamma$ as follows,
\begin{align}
    \gamma =  \sum_{i=1}^\infty  a_s^i \gamma_{i-1} \,,
\end{align}
where $a_s$ is defined in eq.~\eqref{eq:alsexpansion}. 
The cusp anomalous dimension to three-loop order was first extracted from the three-loop non-singlet splitting functions~\cite{Moch:2004pa}, and is given as  
\begin{align}
 \gamma^{\rm cusp}_{0} =& 4 \,, \nonumber
\\
 \gamma^{\rm cusp}_{1} =&  \left(\frac{268}{9}-8 
                 \zeta_2\right) C_A  -\frac{80  T_F N_f}{9}\,, \nonumber
\\
 \gamma^{\rm cusp}_{2} =&\bigg[ \left(\frac{320 \zeta _2}{9}-\frac{224 \zeta _3}{3}-\frac{1672}{27}\right) C_A +\left(64 \zeta _3-\frac{220}{3}\right) C_F\bigg] N_f T_F  \nn
   \\
 +&\left(-\frac{1072 \zeta
   _2}{9}+\frac{88 \zeta _3}{3}+88 \zeta _4+\frac{490}{3}\right) C_A^2  -\frac{64}{27} 
   N_f^2 T_F^2\,.
\end{align}

Finally, the quark and gluon anomalous dimensions of the three-loop order can be extracted from the three-loop quark and gluon form factors~\cite{Moch:2005tm,Gehrmann:2010ue},
\begin{align}
\gamma^q_0 =& -3C_F \, , \nn
\\
\gamma^q_1 =& C_F \left[ C_F \left( -\frac{3}{2} + 12\zeta_2 - 24\zeta_3 \right) + C_A \left( -\frac{961}{54} - 11 \zeta_2 + 26\zeta_3 \right) + T_FN_f \left( \frac{130}{27} +4\zeta_2\right) \right] \,, \nn
\\
\gamma^q_2 = & N_f T_F \bigg[\left(\frac{5188 \zeta _2}{81}-\frac{1928 \zeta _3}{27}+44
   \zeta _4-\frac{17318}{729}\right) C_A C_F \nn 
   \\ &+\left(-\frac{52 \zeta
   _2}{3}+\frac{512 \zeta _3}{9}-\frac{280 \zeta
   _4}{3}+\frac{2953}{27}\right) C_F^2\bigg] +\left(-\frac{80 \zeta _2}{9}-\frac{32 \zeta
   _3}{27}+\frac{9668}{729}\right) C_F N_f^2 T_F^2 \nn
   \\
 &  +\left(-16 \zeta _3 \zeta
   _2+\frac{410 \zeta _2}{3}-\frac{844 \zeta _3}{3}+\frac{494 \zeta
   _4}{3}-120 \zeta _5-\frac{151}{4}\right) C_A C_F^2 \nn 
   \\
  & +\left(-\frac{88}{3}
   \zeta _3 \zeta _2-\frac{7163 \zeta _2}{81}+\frac{3526 \zeta _3}{9}-83
   \zeta _4-136 \zeta _5-\frac{139345}{2916}\right) C_A^2
   C_F\nn
   \\
   &+\left(32 \zeta _3
   \zeta _2-18 \zeta _2-68 \zeta _3-144 \zeta _4+240 \zeta
   _5-\frac{29}{2}\right) C_F^3 \,, \nn
   \\
   \gamma_0^g =& - \beta_0 
    = - \frac{11}{3}\,C_A + \frac43\,T_F n_f \,, \nonumber\\
   \gamma_1^g =& C_A^2 \left( -\frac{692}{27} + \frac{11\zeta_2}{3}
    + 2\zeta_3 \right) 
    + C_A T_F n_f \left( \frac{256}{27} - \frac{4\zeta_2}{3} \right)
    + 4 C_F T_F n_f \,, \nonumber\\
   \gamma_2^g =& C_A^3 \left( - \frac{97186}{729} 
    + \frac{6109\zeta_2}{81} - \frac{319\zeta_4}{3} 
    + \frac{122}{3}\,\zeta_3 - \frac{40}{3} \zeta_2\,\zeta_3 
    - 16\zeta_5 \right) \nonumber\\
   &\mbox{}+ C_A^2 T_F n_f \left( \frac{30715}{729}
    - \frac{2396\zeta_2}{81} + \frac{164\zeta_4}{3} 
    + \frac{712}{27}\,\zeta_3 \right) \nonumber\\
   &\mbox{}+ C_A C_F T_F n_f \left( \frac{2434}{27} 
    - 4 \zeta_2 - \frac{144\zeta_4}{5} 
    - \frac{304}{9}\,\zeta_3 \right) 
    - 2 C_F^2 T_F n_f \nonumber\\
   &\mbox{}+ C_A T_F^2 n_f^2 \left( - \frac{538}{729}
    + \frac{80\zeta_2}{27} - \frac{224}{27}\,\zeta_3 \right) 
    - \frac{44}{9}\,C_F T_F^2 n_f^2 \,,
\end{align}
where the explicit results can also be found in~\cite{Becher:2009qa}.

\section{Instructions of the supplementary material}
We present our results for master integrals valid to transcendentality weight-8 and soft theorem to $\epsilon^2$ in the supplementary material. \texttt{MIsolutions.m} contains the solutions of all 49 three-loop master integrals from the six integral families as defined in section~\ref{sec:integralsFamilies}. \texttt{softTheorem.m} contains the results as shown in eq.~\eqref{eq:b121} to eq.~\eqref{eq:d123}.

\bibliographystyle{JHEP}
\bibliography{softTheorem}




\end{document}